\documentclass[aps,prd,superscriptaddress,twocolumn]{revtex4-1}
\usepackage[parfill]{parskip}    
\usepackage{graphicx}
\usepackage{amssymb}
\usepackage{amsmath}
\usepackage{epstopdf} 
\usepackage{hyperref}
\usepackage{verbatim}
\usepackage{rotating}
\usepackage{graphics, setspace}
\usepackage{multirow}
\usepackage[justification=justified,singlelinecheck=false]{caption}
\usepackage{subcaption}

\usepackage{mathcomp}

\newcommand{\mathsym}[1]{{}}
\newcommand{\unicode}[1]{{}}
\newcommand{\eps}{\epsilon}
\newcommand{\ga}{\gamma}

\newcommand{\half}{{\frac{1}{2}}} 
\newcommand{\om}{\omega} 
\newcommand{\bom}{{\bar\omega}} 
\newcommand{\ep}{\epsilon} 
 
\newcommand{\sinY}{\sin\theta \partial_{\theta} Y_l^m}

\newcommand{\bi}{\begin{itemize}} 
\newcommand{\ei}{\end{itemize}} 
\newcommand{\be}{\begin{equation}} 
\newcommand{\ee}{\end{equation}} 
\newcommand{\beqa}{\begin{eqnarray}} 
\newcommand{\eeqa}{\end{eqnarray}} 
\newcommand{\ba}{\begin{array}} 
\newcommand{\ea}{\end{array}} 
\newcommand{\bea}{\begin{eqnarray}} 
\newcommand{\eea}{\end{eqnarray}} 
\newcommand{\bean}{\begin{eqnarray*}} 
\newcommand{\eean}{\end{eqnarray*}}

\newcommand{\ds}{\displaystyle}

\newcommand{\nn}{\nonumber}

\begin{document}

\author{Ashikuzzaman Idrisy}
\affiliation{Institute for Gravitation and the Cosmos, Center for Particle and Gravitational Astrophysics, Department of Physics, The Pennsylvania State University, University Park, PA 16802, USA}
\author{Benjamin J. Owen}
\affiliation{Institute for Gravitation and the Cosmos, Center for Particle and Gravitational Astrophysics, Department of Physics, The Pennsylvania State University, University Park, PA 16802, USA}
\author{David I. Jones}
\affiliation{Mathematical Sciences, University of Southampton, Southampton SO17 1BJ, UK}

\title{
\emph{R}-mode frequencies of slowly rotating relativistic neutron stars with realistic equations of state}
\date{\today}                                           

\begin{abstract}
The frequencies of \emph{r}-mode oscillations of rotating neutron stars can be
useful for guiding and interpreting gravitational wave and electromagnetic
observations.  The frequencies of slowly rotating, barotropic, and
non-magnetic Newtonian stars are well known, but subject to various
corrections.  After making simple estimates of the relative strengths of these
corrections we conclude that \emph{relativistic} corrections are likely to be the most
important.  For this reason we extend the formalism of K.~H. Lockitch, J.~L.
Friedman, and N. Andersson [Phys.\ Rev.\ D 68, 124010 (2003)], who consider
relativistic polytropes, to the case of \emph{realistic} equations of state.
This formulation results in perturbation equations which are solved using a
spectral method. We find that for realistic equations of state the
\emph{r}-mode frequency ranges from 1.39--1.57 times the spin frequency of the
star when the relativistic compactness parameter  ($M/R$) is varied over the
astrophysically motivated interval 0.11--0.31. The results presented here are
relevant to the design of gravitational wave and electromagnetic \emph{r}-mode
searches, and following a successful \emph{r}-mode detection could help
constrain the high density equation of state. \\ 

\end{abstract}

\maketitle

\section{Introduction}
\label{sec:intro}
\setlength{\parindent}{1cm}  \emph{R}-modes \cite{1978PP} are non-radial
stellar oscillations which can become unstable to gravitational wave (GW)
emission via the CFS \cite{Chandra, FriedmanSchutz78} mechanism
\cite{Ander, FMr-mode}, even in the presence of viscosity
\cite{LindOwenMors98, AndKokSch99}.
This makes \emph{r}-modes a promising source of gravitational waves for ground
based detectors \cite{Bild99, OLCSVA, Andersoon99}.
The energy thus radiated has been used to explain the spins of
newly born neutron stars \cite{LindOwenMors98, AndKokSch99} and of accreting
neutron stars \cite{Bild99, Andersoon99}. The
\emph{r}-modes have also been proposed as a model for quasi-periodic
oscillations of low mass X-ray binaries, and for burst oscillations of
accretion-powered millisecond X-ray pulsars (AMXPs) 
\cite{Watts2012}. Possible detections of \emph{r}-modes in X-ray oscillations have been made from AMXPs XTE J1814\textminus338 \cite{2013r-modeDiscovery} and 4U 1636\textminus536
\cite{r-modeAnotherDiscovery}. (It has been argued
\cite{Andersson2014} that the discovery in \cite{2013r-modeDiscovery} is
inconsistent with the spin-down of the pulsar, though an alternative
explanation has been offered by reference \cite{Lee2014}.)

Because of its physical importance many authors have calculated the \emph{r}-mode frequency. Results are often given in terms of the rotating frame mode angular frequency $\sigma_{\rm R}$, or in dimensionless form as
\be
\label{eq:kappa_definition}
\kappa \equiv \frac{\sigma_{\rm R}}{\Omega} ,
\ee
where $\Omega$ is the rotational angular velocity of the star.  Reference \cite{1978PP} showed that \emph{r}-modes are rotationally restored oscillations and so their frequencies are proportional to the stellar rotation frequency. The authors calculated that for slowly and uniformly rotating Newtonian stars, $\kappa$ is equal to a constant which is independent of the equation of state (EoS):
\be
\label{def:kappa0}
\kappa = \kappa_0= \frac{2m}{l(l+1)},
\ee
where $l$ and $m$ are spherical harmonic indices. It was shown by
\cite{Provost1981} that for barotropic stars the $r$-modes must satisfy $l
=|m|$. The $l=m=2$ \emph{r}-mode, for which $\kappa_0= 2/3$, is the most
susceptible to the CFS instability \cite{LindOwenMors98, AndKokSch99}. 
 
Reference \cite{LindOwen99} extended the slow-rotation expansion for Newtonian stars and found corrections for $\kappa$ to second order in the rotation rate of the star.  Reference \cite{Lock03} examined slowly rotating relativistic stars to leading order in the rotation rate and accounted for metric perturbations. References \cite{Ster, G, YoshidaMain} examined rapidly rotating relativistic stars using the Cowling approximation, in which metric perturbations are neglected. The  rotational and relativistic corrections to $\kappa$ are dependent upon the EoS used to model the star. The studies mentioned thus far used polytropic EoS to simplify the calculation of $\kappa$. 

In this paper we present the first calculation of  $\kappa$ for stellar models constructed from realistic (tabulated) EoS.  We use a subset of the EoS studied by \cite{Read4P, LindIndik}. 
EoS which can not support a maximum mass of least 1.85 solar masses are excluded from our analysis. This mass is a conservative upper limit derived from the 99.7\% confidence limit of the observed ``1.97 $M_{\odot}$''pulsar, see Fig.~2 of \cite{Demorest} for more details. This left 14 EoS for which a range of $\kappa$ values were calculated over a range of masses.  

There are several  applications of our results. Our calculation of $\kappa$
can be used to interpret electromagnetic observations of \emph{r}-modes, such as
those (possibly) of \cite{2013r-modeDiscovery, r-modeAnotherDiscovery}. Our calculation can also be used in collaborative work between GW detectors and (electromagnetic) astronomical
observatories. Assuming \emph{r}-mode GWs are detected from a previously unknown pulsar, our range of $\kappa$
would give electromagnetic astronomers a frequency band in which to search for pulsations from
rotation. If one has the \emph{r}-mode frequency (from GW data) and the pulsation
frequency (from electromagnetic data), it is possible to get the pulsar's
compactness which might be used to constrain the EoS. Finally, for GW
searches conducted on pulsars with known spin frequencies, such as the Crab,
our results define a narrow frequency band over which to search for
GWs from \emph{r}-modes. A narrow band search \cite{Crab2008}
has already been conducted on the Crab but it was not looking for gravitational
waves from \emph{r}-modes. Rather the search was centered around the usual, two times the spin-frequency of the pulsar. Our results can also be used by a new
narrow-band search pipeline \cite{narrowband} which claims to be twice as
sensitive as the previous Crab search.

The outline of the paper is as follows. In Sec.~\ref{sect:effects} we estimate how general relativity affects the \emph{r}-mode frequency in comparison to other physical phenomena such as the star's crust, rotation rate, magnetic fields, and stratification. In Sec.~\ref{Formulation} we present the  formulation of the  \emph{r}-mode oscillation problem found in \cite{Lock03}. In Sec.~\ref{NumSol} we discuss the numerical methods used to solve the equations that arise from this formulation. We also give more details than \cite{Lock03}, including convergence details for our code, and accuracy estimates for our results. In Sec.~\ref{Results} we discuss the results of applying our numerical solution to both polytropic and realistic equations of state, with a focus on the latter. Finally in Sec.~\ref{Conclusions} we draw conclusions from our results and examine the aforementioned applications in light of these results. Throughout we use geometrized units, where Newton's gravitational constant and the speed of light are unity.

\section{Physical phenomena which affect the \emph{r}-mode frequency}
\label{sect:effects}

We now estimate the importance of various corrections to the Newtonian slow-rotation \emph{r}-mode frequency.  In common with much of the literature, we make use of the dimensionless rotating frame mode frequency $\kappa$ defined in Eq.~(\ref{eq:kappa_definition}).  Note that the corresponding gravitational wave emission will be at the inertial frame mode frequency $\sigma_{\rm I}$, related to the rotating frame mode frequency by:
\begin{equation}
\label{eq:kappa_def}
\sigma_{\rm I} = (\kappa - m) \Omega = \sigma_{R} - m \Omega. 
\end{equation}
For the $l=m=2$ \emph{r}-mode of a slowly rotating Newtonian star,
$\sigma_{\rm R} = 2\Omega/3$ while $|\sigma_{\rm I}| = 4\Omega/3$.  Strictly,
$\sigma_{\rm I} = -4\Omega/3$; the negative sign is a consequence of the
opposite sense of rotation of the patterns produced by the mode as viewed in
the inertial and rotating frames. This opposite sense of rotation is
responsible for the CFS instability. It follows that a decrease of $\kappa$ by
a small fraction will increase the inertial frame gravitational wave
frequency by half of that fraction. This must be borne in mind when designing
gravitational wave searches (see Sec.~\ref{Conclusions} for some discussion of
detection issues).

\subsection{General relativity} \label{sect:GR}

The importance of general relativistic effects can be estimated by looking at the ratio of stellar mass to radius, $M/R$, a dimensionless measure of the compactness of the star:
\begin{equation}
\label{eq:M_over_R}
\frac{M}{R} \approx 0.207 \left(\frac{M}{1.4 M_\odot}\right) \left(\frac{10 \, \rm km}{R}\right).
\end{equation}
It follows that departures from the Newtonian $\kappa = \kappa_0$ results, at the level of a few tens of percent, can be expected when relativistic effects are included. This expectation is confirmed by the post-Newtonian and fully relativistic numerical calculations of \cite{Lock03}, who considered polytropic stars.  For instance, for a uniform density star with $M/R = 0.207$, they found that $\kappa / \kappa_0 \approx 0.85$, a reduction of $\sim 15\%$ compared to the Newtonian case.  Note that (consistent with the results to be presented in this paper),  \cite{Lock03} found that the effect of relativity is to decrease the rotating frame mode frequency, and therefore increase the inertial frame frequency.

Given that we will conclude that relativistic effects are likely to be the
dominant factor influencing the \emph{r}-mode frequency, it is worth
considering how large a spread of the compactness parameter might be found in
the neutron star population.  This is of immediate astrophysical interest, as
it would determine how large a range in gravitational wave frequency must be
searched when looking for \emph{r}-mode emission from a pulsar of known spin
frequency (but unknown compactness).  In the numerical calculations that
follow, for each chosen realistic EoS, a range of masses from 1 $M_\odot$ up
to a value close to the maximum mass (specific to that equation of state) are
considered.  The lower limit of 1 $M_\odot$ was adopted since the
lepton-rich, hot matter in supernova explosions most likely does not support
proto-neutron stars with smaller masses.
Also, most measured masses with tight error bars are greater than this; see
Fig.~1 of \cite{lp10}. Thus we have taken the conservative lower limit of 1
$M_\odot$. 

This led to a range of compactnesses $0.110 \leq  M/R \leq 0.310$. The graphs
of \emph{r}-mode frequency that follow are plotted over this range.  That this
is a sensible estimate of the range of possible compactnesses of realistic
neutron stars can be confirmed from Fig.~2 of \cite{lp07}, which shows
mass-radius curves for a large collection of realistic EoS, some of which are
also considered in this paper.  As illustrated, there is a hard upper limit of
$M/R \lesssim 0.350$ that comes from the constraint that the EoS be causal.
The maximum mass members of soft EoS come close to (but do not quite reach)
this limit, e.g.\  the EoS AP4, which is one of the EoS considered here.  In
terms of a lower limit on compactness, low mass stars with stiff equations of
state are relevant.  Reference \cite{lp07} find EoS whose 1 $M_\odot$ members
have $R \approx 14.5$ km, corresponding to $M/R \approx 0.103$. However, for
the EoS considered here this lower limit would result in stars with masses
less than 1 $M_\odot$. Therefore we increase this lower limit to $0.110$.
Taken together, we see that the range of compactnesses considered  in this
paper includes neutron stars presently considered realistic in the
literature.  We will return to this in Sec.~\ref{Conclusions}, after having calculated the range in \emph{r}-mode frequencies corresponding to this range in compactness.

\subsection{Rapid rotation} \label{sect:rotation}

The effect of stellar rotation on the \emph{r}-mode frequency was considered by  \cite{LindOwen99}, who calculated the leading order correction to the mode frequency, as quantified by a parameter $\kappa_2$ satisfying 
\be
\label{def:kappa2}
\kappa = \kappa_0 +\kappa_2  \frac{\Omega^2}{\pi G \bar{\rho}_0}. 
\ee   
Here $\bar{\rho}_0$ is the average mass density of the corresponding non-rotating star and $\kappa_2$ is  dimensionless, of order unity,  and dependent upon the  equation-of-state.  The factor
\begin{equation}
\label{eq:Omega_squared_over_rho}
\frac{\Omega^2}{\pi G \bar\rho_0} = 0.145 \, \left(\frac{f_{\rm spin}}{716 \, \rm Hz}\right)^2
\left(\frac{R}{10^6 \, \rm cm}\right)^3 \left(\frac{1.4 M_\odot}{M}\right)
\end{equation}
is a dimensionless measure of the effect of rotation on the star. We have
scaled the spin frequency $f_{\rm spin} = \Omega / 2\pi$ to a value of $716$
Hz, the spin rate of the fastest observed millisecond pulsar \cite{ATNF}.
Taking a representative value of $\kappa_2 \approx 0.29$ from
\cite{LindOwen99}, we see that rapid rotation can increase the value of
$\kappa$ by $\sim 6\%$ for the fastest rotating stars,  while rotational
corrections rapidly become negligible for more slowly spinning stars.   It follows that rotational effects can indeed be significant, but probably never dominate the relativistic ones.  Note that the sign of the frequency shift corresponds to a decrease in the gravitational wave frequency, and so acts oppositely to the relativistic effects described above.

\subsection{The crust} \label{sect:crust}

The presence of a solid crust is very important for \emph{r}-mode damping \cite{BilUsh2000, LevUsh2001},  and it can have an effect on the mode frequency as well.  In addition to Coriolis restoring forces acting throughout the star, there are also elastic restoring forces in the crust.  Information on how this influences the \emph{r}-mode frequency can be extracted from Fig.~1 of \cite{LevUsh2001}. For sufficiently slow rotation rates, the mode frequency is close to the standard $\kappa = \kappa_0 = 2/3$ result, with the fluid core but not the solid crust participating in the motion.  For sufficiently high rotation rates the mode frequency is again close to $\kappa = \kappa_0 =  2/3$, but now the whole star, crust plus core, participates in the motion.  For intermediate spin rates, there is an avoided  crossing, which means that  the `\emph{r}-mode' is more accurately described as a hybrid rotational--elastic mode.  

From Fig.~1 of \cite{LevUsh2001} it seems that the departure from the $\kappa = \kappa_0 = 2/3$ result is significant (i.e.\ more than a few percent and can be discerned by eye) over the spin frequency interval $0.05 \lesssim \Omega / \Omega_{\rm K} \lesssim 0.1$, where $\Omega_{\rm K}$ is the  Keplerian angular velocity of the star.  Being dependent upon the EoS, this quantity is not known accurately, but taking a representative value of  $\Omega_{\rm K}/(2\pi) \sim 1500$ Hz, this corresponds to the spin interval $75 < f_{\rm spin} / {\rm Hz} < 150$, so crustal corrections could be relevant for some milli-second pulsars.

Looking at the right hand panel of Fig.~1 of \cite{LevUsh2001}, we see that departures from $\kappa = \kappa_0 = 2/3$ of $\sim \pm 20\%$ are possible. This is comparable with the  shift of Sec.~\ref{sect:GR}, but is double-sided, i.e.\ the mode frequency might be shifted up or down.  However, the modification of the mode frequency at this level only applies over a narrow range in spin frequency so it is unlikely to affect most stars.

\subsection{Other effects} \label{sect:other_effects}

There will be other factors that will have effects on the \emph{r}-mode frequency.  We very briefly mention two more here.

Real neutron stars are stratified, with radial entropy and composition gradients. The effect of stratification was considered by \cite{YosLee2000}, who found that while the majority of the inertial modes are significantly affected by stratification, the nodeless $l=m$ \emph{r}-modes are relatively unaffected; see Fig.~4 of \cite{YosLee2000}. The effect of stratification on inertial modes was also investigated by \cite{PasHas2009}, who found the \emph{r}-mode frequency was affected only very slightly.  This was shown to be true even for very rapidly rotating stars; see Fig.~12 of \cite{PasHas2009}.

Magnetic fields will also alter the \emph{r}-mode frequency, but the effect is again likely to be  slight, see \cite{HoLai2000, MorRez2002}, or the numerical simulations of \cite{LanJones2010}. Physically, the smallness of the corrections corresponds to the magnetic restoring  forces being small compared the Coriolis restoring forces.  It should, however, be pointed out that the above references consider non-superconducting stars. The effect of superconductivity may make magnetic corrections more important, but quantitative estimates of such effects are not currently available.

\subsection{Summary}

We have presented estimates of the importance of various effects on the \emph{r}-mode spin frequency, using  some simple estimates and results from the literature.  General relativistic effects can have a significant influence on the \emph{r}-mode frequency, at the level of tens of percent.  The effects of rapid rotation are insignificant in all but the fastest spinning pulsars. The effects of an elastic crust are slightly more difficult to quantity, but it seems likely they will only be competitive with relativistic ones in rather narrow intervals in stellar spin frequency, and so are unlikely to be significant in the majority of the known pulsars.

The likely dominance of relativistic effects motivates the careful treatment of relativistic stars with \emph{realistic} EoS presented in the remainder of this paper.

\section{Formulation}
\label{Formulation}
In this section we summarize the formulation of the \emph{r}-mode oscillation problem in order to provide context, and establish terminology and notation. The full details are available in \cite{Lock01, Lock03}. In this formulation, the PDEs resulting from the perturbed Einstein equations for a perfect fluid star, are turned into ODEs via spherical harmonic expansion. This expansion is only possible if the star is assumed to be slowly rotating.   


\subsection{Equilibrium Solution}
\label{sec:equilSol}
In order to derive the perturbation equations, equilibrium solutions of slowly rotating and non-rotating stars must be found first. 
\subsubsection{Equilibrium non-rotating star}
The non-rotating equilibrium solution is found by solving the Einstein equations $G_{\alpha \beta} = 8\pi T_{\alpha \beta}$, where $G_{\alpha \beta}$ is derived from the line element: 
\be
ds^2 = -e^{2\nu(r)} dt^2 + e^{2\lambda(r)} dr^2 + r^2 d \theta^2 
	+ r^2 sin^2 \theta d \varphi^2,
\ee
$T_{\alpha \beta}$ is the energy-momentum tensor for a perfect fluid:
\be
T_{\alpha\beta} = (\epsilon+p)u_\alpha u_\beta + p g_{\alpha\beta},
\ee
$\ep(r)$ is fluid energy density, $p(r)$ is fluid pressure, and 
\be
u^\alpha = e^{- \nu} t^\alpha
\ee
is the fluid 4-velocity with $t^\alpha=(\partial_t)^\alpha$ the time-like Killing vector. Applying this information and comparing $G_{\alpha \beta} = 8\pi T_{\alpha \beta}$ term-by-term leads to the Oppenheimer-Volkov (OV) equations  \cite{carroll2004}. Therefore solving the OV equations is equivalent to solving the Einstein equations. 

The OV equations must be solved numerically for most equations of state. The numerical solution is better realized when one uses the enthalpy, $h$, of the star instead of the radial distance, $r$, as the dependent variable  \cite{Lindblom1992} (we refer to these as the OVL equations). The OVL equations for a non-rotating star are 
\be
\frac{dr}{dh} = -\frac{r(r-2M)}{(M+4\pi r^3p)},
\label{drdh}
\ee
and
\be
\frac{dM}{dh} = 4\pi r^2\ep \frac{dr}{dh} ,
\label{dMdh}
\ee
where $M(h=0)$ is the mass of the star. The metric functions $\lambda$ and $\nu$ are found using
\be
\nu(h)-\nu_c = h_c - h,
\label{nueq}
\ee
where 
\be
\nu_c = - h_c + \frac{1}{2} \ln \left ( 1 + \frac {2 M(h=0)}{r(h=0)} \right )
\ee 
and
\be
\lambda(h) = -\frac{1}{2} \log \left[1 - \frac{2 M(h)}{r(h)} \right].
\ee

Just like the OV equations the OVL equations are singular at the center of the star.  Therefore, the numerical integration is started near the center, $h=h_c$, using the following truncated power series solutions
\bea
r(h) = \left[ \frac{3(h_c-h)}{2\pi (\ep_c+3p_c)}\right]^\half
\bigg \{1 - \frac{1}{4}\left[\ep_c-3p_c+\frac{3}{5}\ep_1\right] \nonumber \\
\times \frac{(h_c-h)}{(\ep_c+3p_c)} \biggr \},
\eea
\be
M(h) = \frac{4\pi}{3}\ep_cr^3(h)\left\{
1 + \frac{3\ep_1}{5\ep_c}(h_c-h)\right\},
\ee
where $\ep_c$ is the central energy density, $p_c$ is the central pressure and 
\be
\ep_1 = - \left. \frac{d\ep}{dh}\right|_{h=h_c} . 
\ee
The integration is carried out to $h=0$, which is guaranteed to be the surface of the star. 

In order to solve the OVL equations one also needs to specify an equation of state. The details of using polytropic EoS in the OVL equations are explained in \cite{LockThesis}. Realistic EoS are presented as tables with columns given by values of pressure, $p_i$, energy density, $\epsilon_i$, and baryon number density $n_i$, where the $i$ subscript indexes the row of the table. 
The values in the columns must be interpolated in order to get a well-behaved
EoS which can be used with an OV or OVL solver. We use the  interpolation scheme of \cite{LindIndik} for our analysis. 

The interpolation scheme of \cite{LindIndik}  assumes a power law relationship between pressure and energy density, 
\be
\frac{p}{p_i} = \left(\frac{\epsilon}{\epsilon_i}\right)^{c_{i+1}},
\label{e:piInterpolate}
\ee
where
\be
c_{i+1}=\frac{\log(p_{i+1}/p_i)}{\log(\epsilon_{i+1}/\epsilon_i)},
\ee
for $\epsilon_i \leq \epsilon \leq \epsilon_{i+1}$.  
Using this and the definition of the co-moving enthalpy 
\be
h(p)=\int_0^p\frac{dp'}{\epsilon(p')+p'},
\label{e:AppendixEnthalpy}
\ee
a column of values, $h_i=h(p_i)$, is generated and is used to get the piecewise function 
\begin{eqnarray}
&&\!\!\!\!\!\epsilon(h)=\nonumber\\
&&\!\!\!\!\! \epsilon_i\left\{\frac{\epsilon_i+p_i}{p_i}
\exp\left[\frac{c_{i+1}-1}{c_{i+1}} (h-h_i)\right]-\frac{\epsilon_i}{p_i}
\right\}^{1/(c_{i+1}-1)}\\
\end{eqnarray}
for $h_i\leq h \leq h_{i+1}$.~Using Eq.~\eqref{e:piInterpolate} one can get $p(h)$. 

The interpolation scheme of \cite{HP} by contrast assumes a power law relationship between pressure and the number density
\begin{equation}
\label{hanselinterp}
p(n) = p_i \left (\frac{n}{n_i} \right )^{\ga_i} ,
\end{equation}
where 
\be
\ga_i = \frac{\ln p_{i+1} - \ln p_i}{\ln n_{i+1} - \ln n_{i}},
\ee
for $n_i \leq n \leq n_{i+1}$.~This scheme requires creating an auxiliary column of values $\tilde \epsilon_i$, 
\be
\tilde \eps_{i+1} = \tilde \eps_i  + \frac{1}{\ga_i -1} \left (\frac{ p_{i+1}}{n_{i+1}} - \frac{p_i}{n_i} \right)
\ee
to get the auxiliary energy density $\tilde \epsilon(n)$, \be
\tilde \eps(n) = \tilde \eps_i + \frac{1}{\ga_i - 1}\left(\frac{p}{n} - \frac{p_i}{n_i} \right)
\ee
which is used to find the energy density
\be
\epsilon (n) = n \left [ \tilde \eps(n) + m_n \right], 
\ee
where $m_n$ is the mass of a neutron. For our purposes we require $\epsilon(h)$, and $p(h)$. Thus we use Eq.~\eqref{e:AppendixEnthalpy} to find a column of values for $h_i$. Then we interpolate the $h_i$ and $n_i$ values using cubic splines, to define the function $n(h)$, which we substitute into $\epsilon(n)$, and $p(n)$. 

Finally, in our implementation of a simple spline interpolation (see \cite{NumRec} for details) we assume a power law relationship between pressure and energy density. Whereas  the interpolation schemes of \cite{LindIndik,HP} take the first law of thermodynamics into account this alternative spline scheme does not. Therefore this scheme mainly serves as a test of how much the first law affects the result for $\kappa$. 
 
We used a linear interpolation of $\log p_i$ and $\log \epsilon_i$ values to determine the power law between the points. Making use of Eq. \eqref{e:AppendixEnthalpy} to get values for $h_i$, we interpolated $h_i$ with $p_i$ and $\epsilon_i$ using a quadratic spline, to get the function $p(h)$ and $\epsilon(h)$ respectively. The quadratic spline was used because first order splines lead to discontinuities in the solutions to the OVL equations, and the third order splines lead to extra inflection points in the solutions.

To check that our results are robust to the interpolation method used, we compared the scheme in \cite{LindIndik} to that found in \cite{HP}, and our own spline interpolation for a sample group of EoS. We found that the percent difference in $\kappa$ from the different schemes was less than $0.3\%$ overall, and in some cases less than $0.1\%$.  From these small percent differences, we see that our code is robust to the interpolation scheme used.
\subsubsection{Equilibrium rotating Star}
The equilibrium solution for a rotating star is again found by solving the Einstein equations. This time, $G_{\alpha \beta} $ is derived from the line element:
\bea
\label{equil_metric}
ds^2 = -e^{2\nu(r)} dt^2 &+& e^{2\lambda(r)} dr^2 + r^2 d \theta^2  \\ \nonumber 
 &+& r^2 \sin^2 \theta d \varphi^2 - 2 \omega(r) r^2 \sin^2\theta\, dt\, d\varphi 
\eea
with the definition
\be
\bom(r)\equiv\Omega-\om,
\ee
where $\omega(r)$ accounts for the frame-dragging effect. The line element in Eq.~\eqref{equil_metric} is only correct up to order $\Omega$. This slow rotation limit means that the star retains its spherical geometry, since the centrifugal deformation of its figure is an order $\Omega^2$ effect \cite{Hartle67}.  In the rotating case the fluid 4-velocity becomes  
\be
u^\alpha = e^{-\nu} ( t^\alpha + \Omega \varphi^\alpha)
\label{equil_4v}
\ee
where  $t^\alpha=(\partial_t)^\alpha$ and $\varphi^\alpha=(\partial_\varphi)^\alpha$, are respectively the time-like and rotational Killing vectors. For a rotating star the equilibrium solution comes from solving the OVL Eqs.~\eqref{drdh}--\eqref{dMdh} and the Hartle equation \cite{Hartle67}. In the enthalpy formulation, which was extended to slow rotation by \cite{Read4P}, the Hartle equation is broken up into a pair of first order ODEs as
\be
\frac{d\bom}{dh} = e^{(\nu-\nu_c+\lambda)} f \frac{dr}{dh}
\label{dbomdh}
\ee
and
\be
\frac{df}{dh} = \left[ 16\pi(\ep+p)e^{-(\nu-\nu_c-\lambda)}\bom
-\frac{4}{r}f\right] \frac{dr}{dh}.
\label{dfdh}
\ee
These equations are also singular at the center of the star, and so the following power series are used to begin the integration,
\be
\bom(h) = \om_c \left\{
1 + \frac{12(\ep_c+p_c)}{5(\ep_c+3p_c)}(h_c-h)\right\},
\ee
\bea
f(h) =  \frac{16\pi}{5}(\ep_c+p_c)\om_cr(h) \biggr \{ 1 &+&  \frac{5}{7}\biggr[\frac{6(2\ep_c-3p_c)}{5(\ep_c+3p_c)}  \nonumber \\
&+&\frac{\ep_1}{(\ep_c+p_c)}\biggr]  (h_c-h) \biggr \} \nonumber \\
\eea
\normalsize
After solving Eqs.~\eqref{drdh}--\eqref{dMdh} , and Eqs.~\eqref{dbomdh}--\eqref{dfdh} we numerically invert the solution for Eq.~\eqref{drdh}. Using this we change the functions $\lambda(h), \nu(h), \ldots$ into  $\lambda(r), \nu(r)$, etc. 

\subsection{Perturbation equations} 
\setlength{\parindent}{1cm} In this section we sketch out the derivation of the perturbation equations, full details of which are in \cite{Lock03}. 
The metric and fluid perturbation terms in these equations are expanded in terms of  scalar ($Y_l^m$), vector ($r \nabla Y_l^m$, $r \times \nabla Y_l^m$) and tensor ($\nabla_{\beta} Y_l^m u_\gamma$) spherical harmonics. This basis makes it possible to classify perturbations as axial or polar parity. Axial parity modes have the same parity as $r \times \nabla Y_l^m$, whereas polar parity modes have the parity of $Y_l^m$ and $\nabla Y_l^m$. The generic oscillation will be a combination of polar and axial modes. However, the leading order term in the expansion will either be of polar parity or axial parity.  This leads to the terminology of axial-led and polar-led modes, with \emph{r}-modes being the former. This classification works for rotating as well as non-rotating stars because the parity of the leading order term does not change due to rotation. 

\subsubsection{Perturbations of non-rotating stars}
The equilibrium  configuration for a non-rotating star is given as a solution to the OVL Eqs.~\eqref{drdh}--\eqref{dMdh}. The perturbations are
\begin{eqnarray}
\delta\ep &=& \delta\ep(r) Y_l^m, \\
\delta p &=& \delta p(r) Y_l^m, \\
\delta u_P^{\alpha} &=& \biggl\{ \half H_0(r) Y_l^m t^\alpha + \frac{1}{r}
W(r) Y_l^m r^{\alpha} + V(r) \nabla^{\alpha} Y_l^m  \biggr\}
\nonumber\\
& & \times e^{-\nu},\\
\delta u_A^{\alpha} &=& 
- U(r) e^{(\lambda-\nu)} \epsilon^{\alpha\beta\gamma\delta} 
\nabla_{\beta} Y_l^m u_\gamma \nabla_{\delta}\, r.
\end{eqnarray}  \\
Notice  that $\delta u_P^{\alpha}$ is of polar parity and $\delta u_A^{\alpha}$ is of axial parity. Employing the Regge-Wheeler gauge and expanding the metric in tensor spherical harmonics, the metric perturbation of polar-parity mode can be written
\begin{widetext}
\be
h_{\mu\nu}^P = \left[
\ba[c]{cccc}
H_0(r) e^{2\nu} & H_1(r) & 0 & 0 \\
H_1(r) & H_2(r) e^{2\lambda} & 0 & 0 \\
0 & 0 & r^2 K(r) & 0\\
0 & 0 & 0 & r^2 \sin^2\theta K(r)
\ea
\right] Y_l^m,
\label{GR:h_po}
\ee
and that of an axial-parity mode can be written
\be
h_{\mu\nu}^A = \left[
\ba[c]{cccc}
0 & 0 & -h_0(r) \csc\theta\partial_\varphi Y_l^m & h_0(r)\,\sinY\\
0 & 0 & -h_1(r) \csc\theta\partial_\varphi Y_l^m & h_1(r)\,\sinY\\
\mbox{ symm} & \mbox{ symm} & 0 & 0 \\
\mbox{ symm} & \mbox{symm} & 0 & 0 \\
\ea
\right] 
\label{GR:h_ax}
\ee
where ``symm'' indicates components obtained by symmetry.
\end{widetext}

The final step in finding the perturbation equations is to examine $\delta G_{\alpha \beta} = 8 \pi \delta T_{\alpha \beta}$ term-by-term. This leads to 10 differential equations for $(H_0, H_1,H_2, K, h_0, W, V, U , \delta \epsilon, \delta p )$. The equations decouple into equations for $( H_1, h_0, W, V, U)$ and $(H_0,H_2, K,\delta \epsilon, \delta p)$. Under the assumption of linear stability, reference \cite{LockThesis} showed that for non-radial oscillations $H_0 = H_2 = K = \delta \epsilon= \delta p= 0$. 
Thus the perturbation equations of $\mathcal{O}(1)$ in $\Omega$ are
\be
V_l [ l(l+1)(\epsilon+p) ]   - e^{-(\nu+\lambda)}  
\left[(\epsilon+p)e^{\nu+\lambda} r W_l \right]' = 0 ,
\label{GR:sph_V} 
\ee
\begin{widetext}
 \be
r^2 h_l^{''} - r^2 (\nu'+\lambda') h_l'  + \biggl[ (2-l^2-l) r^2 e^{2\lambda} -  r (\nu'+\lambda')  - 2 \biggr] h_l  -  4r (\nu'+\lambda') U_l  \ = 0.  
\label{GR:sph_h_0''} 
\ee
\end{widetext} 
In the Newtonian limit Eq.~\eqref{GR:sph_V} corresponds to conservation of mass and Eq.~\eqref{GR:sph_h_0''}, which relates the metric perturbation to the fluid perturbation, reduces to identity (vanishing metric perturbation). We have made slight algebraic changes to Eq.~\eqref{GR:sph_V}  and Eq.~\eqref{GR:sph_h_0''} from the way they appear in \cite{Lock03}, so that they are easier to use in the numerical computation discussed in Sec.~\ref{NumSol}. 

\subsubsection{Perturbations of slowly rotating stars}
Similar to the non-rotating case the fluid perturbation is decomposed into spherical scalar and vector harmonics. But this time the Lagrangian perturbation formalism is used. In general the Lagrangian change of a quantity Q, $\Delta Q$, is related to the Eulerian change, $\delta Q$, via $\Delta Q = \delta Q + {\cal{L}_{\xi}} Q$, where $\cal{L}_{\xi}$ is the Lie derivative with respect to the fluid displacement vector ${\xi}$. Here, the displacement vector is defined as:
\begin{widetext}
%
\be
\xi^{\alpha} \equiv \frac{1}{i\kappa\Omega} \sum_{l=m}^\infty 
\biggl\{  \ba[t]{l}
\ds{\frac{1}{r} W_l(r) Y_l^m r^{\alpha} 
+ V_l(r) \nabla^{\alpha} Y_l^m } - \ds{i U_l(r) P^\alpha_{\ \mu} \epsilon^{\mu\beta\gamma\delta} 
\nabla_{\beta} Y_l^m \nabla_{\!\gamma} \, t \nabla_{\!\delta} \, r}
\biggr\} e^{i\sigma t},
\ea
\label{xi_exp}
\ee
where
\be
P^\alpha_{\ \mu} \equiv e^{(\nu+\lambda)} 
\left( \delta^\alpha_{\ \mu}
- t_\mu \nabla^\alpha t
\right).
\ee
It should be noted that $W_l, V_l$ are of polar parity whereas $U_l$ is of axial parity. Again using the Regge-Wheeler gauge the metric perturbation is
\be
h_{\mu\nu} = \sum_{l=m}^\infty \left[
\ba[c]{cccc}
H_{0,l}(r)e^{2\nu}Y_l^m & H_{1,l}(r)Y_l^m & 
h_{0,l}(r) \,(\frac{m}{\sin\theta})Y_l^m & i h_{0,l}(r)\,\sinY\\
H_{1,l}(r)Y_l^m & H_{2,l}(r)e^{2\lambda}Y_l^m & 
h_{1,l}(r) \,(\frac{m}{\sin\theta})Y_l^m & i h_{1,l}(r)\,\sinY\\
\mbox{symm} & \mbox{symm} &
r^2K_l(r)Y_l^m & 0\\
\mbox{symm} & \mbox{ symm} &
0 & r^2sin^2\theta K_l(r)Y_l^m
\ea
\right]e^{i\sigma t},
\label{h_components}
\ee
where the polar parity components are $H_{0,l}, H_{1,l}, K$, and the axial parity components are $h_{0,l}, h_{1,1}$. The coefficients can be grouped as:
\begin{eqnarray}
W_l, V_l, U_l, H_{1,l}, h_{0,l} \sim \mathcal{O}(1), \\
H_{0,l}, H_{2,l},K_l h_{1,l}, \delta \epsilon, \delta p \sim \mathcal{O}(\Omega) .
\end{eqnarray} 
The $\mathcal{O}(1)$ coefficients obey the $\mathcal{O}$(1) Eqs.~\eqref{GR:sph_V}--\eqref{GR:sph_h_0''}. 

The definition of $\xi^{\alpha}$ leads to $\kappa \Omega$ terms in the
perturbation equations, and so only the $\mathcal{O}(1)$ variables are kept in
the perturbation equations. This ensures that the order of the equations is no higher than $\mathcal{O}(\Omega)$. Reference \cite{Lock01} derives the $\mathcal{O}(\Omega)$ equations by invoking the conservation of circulation for an isentropic fluid, which gives 
\bea
\left[ l(l+1)\kappa\Omega(h_l+U_l)-2m{\bar\omega}U_l\right] + (l+1) &Q_l& \left[ \frac{e^{2\nu}}{r} \partial_r \left(r^2{\bar\omega}e^{-2\nu}\right)W_{l-1} - 2(l-1){\bar\omega}V_{l-1}  \right]  \nonumber \\ 
&-& \, l Q_{l+1} \left[\frac{e^{2\nu}}{r}\partial_r \left(r^2{\bar\omega}e^{-2\nu}\right)W_{l+1} +2(l+2){\bar\omega}V_{l+1} \right]  = 0,
\label{om_th_ph} 
\eea \\
%
%
%
%
\bea 
&& (l-2)Q_{l-1}Q_l \left[ 
-2\partial_r\left({\bar\omega}e^{-2\nu}U_{l-2}\right) 
+\frac{(l-1)}{r^2}\partial_r\left(r^2{\bar\omega}e^{-2\nu}\right)U_{l-2} 
\right]  \nn \\ 
&&\nn \\ 
&&+ Q_l \biggl[ 
	\ba[t]{l} 
	(l-1)\kappa\Omega\partial_r\left(e^{-2\nu}V_{l-1}\right) 
	-2m\partial_r\left({\bar\omega}e^{-2\nu}V_{l-1}\right) \\ 
	 \\ 
	+\frac{m(l-1)}{r^2}\partial_r\left(r^2{\bar\omega}e^{-2\nu}\right)V_{l-1} 
	+(l-1)\kappa\Omega e^{-2\nu}\left(\frac{16\pi r(\epsilon+p)}{(l-1)l} 
					-\frac{1}{r}\right)e^{2\lambda}W_{l-1} 
	\biggr] 
	\ea \nn \\ 
&&\nn \\ 
&&+\biggl[ 
	\ba[t]{l} 
	m\kappa\Omega\partial_r\left[e^{-2\nu}(h_l+U_l)\right] 
	+2\partial_r\left({\bar\omega}e^{-2\nu}U_l\right) 
	\left((l+1)Q_l^2-l Q_{l+1}^2\right) \\ 
	 \\ 
	+\frac{1}{r^2}\partial_r\left(r^2{\bar\omega}e^{-2\nu}\right)U_l 
	\left[m^2+l(l+1)\left(Q_{l+1}^2+Q_l^2-1\right)\right] 
	\biggr] 
	\ea \nn \\ 
&&\nn \\ 
&&- Q_{l+1} \biggl[ 
	\ba[t]{l} 
	(l+2)\kappa\Omega\partial_r\left(e^{-2\nu}V_{l+1}\right) 
	+2m\partial_r\left({\bar\omega}e^{-2\nu}V_{l+1}\right) \\ 
	 \\ 
	+\frac{m(l+2)}{r^2}\partial_r\left(r^2{\bar\omega}e^{-2\nu}\right)V_{l+1} 
	+(l+2)\kappa\Omega e^{-2\nu}\left(\frac{16\pi r(\epsilon+p)}{(l+1)(l+2)} 
					-\frac{1}{r}\right)e^{2\lambda}W_{l+1} 
	\biggr] 
	\ea \nn \\ 
&&\nn \\ 
&&+(l+3)Q_{l+1}Q_{l+2} \left[ 
2\partial_r\left({\bar\omega}e^{-2\nu}U_{l+2}\right) 
+\frac{(l+2)}{r^2}\partial_r\left(r^2{\bar\omega}e^{-2\nu}\right)U_{l+2} 
\right] = 0 ,
\label{om_r_th}  
\eea
%
%
\end{widetext}
%
where the constants $Q_l$ are defined as 
\be 
Q_l \equiv \left[ 
\frac{(l+m)(l-m)}{(2l-1)(2l+1)} 
\right]^{1/2}. 
\ee 
\subsection{Boundary Conditions}
In order to solve Eqs.~\eqref{GR:sph_V}--\eqref{GR:sph_h_0''} and
Eqs.~\eqref{om_th_ph}--\eqref{om_r_th} we need to apply the appropriate
boundary conditions. Notice that the perturbation equations are a set of
linear ODEs. This indicates that multiplying a solution by a constant gives
another solution, which means that the boundary conditions must take the form
$\zeta(h_l,U_l, V_l, W_l) = 0$, where $\zeta$ represents an arbitrary linear
combination. Alternatively, the boundary condition must be given in terms of a
condition on a logarithmic derivative, e.g. $U_l'/U_l = constant$.

First let us consider the boundary conditions near the center of the star. These are also known as the regularity conditions:
\bea
U_l (r \rightarrow 0 ) &=& \left( \frac{r}{R} \right)^{l}\overline{U_l}(r), \hspace{5mm} W_l (r\rightarrow 0) =\left( \frac{r}{R} \right)^{l+1}\overline{W_l}(r) \nonumber \\
h_l(r \rightarrow 0) &=&  \left( \frac{r}{R} \right)^{l} \overline{h_l}(r), \hspace{5mm} V_l (r\rightarrow 0) = \left( \frac{r}{R} \right)^{l+1}\overline{V_l}(r) \nonumber \\
& & 
\label{reg_cond}
\eea
where $R$ is the surface of the star, and the barred functions are slowly varying. Only two of these boundary conditions are linearly independent as shown by \cite{Lock03}.

Next let us examine the boundary conditions at the surface of the star. The Lagrangian perturbation of the pressure is zero at the surface, which leads to%
\be
W_l(R) = 0. 
\label{eq:BConW}
\ee
Note that $h_l$ is the only unknown function defined outside of the star, where it obeys 
\be 
\left(1-\frac{2M}{r} \right) \frac{d^2 h_l}{dr^2} - \left[ \frac{l(l+1)}{r^2}  
	- \frac{4M}{r^3} \right] h_l  = 0, 
\label{h_l''_ext} 
\ee 
which has the exact solution
\be 
h_l(r) = \sum_{s=0}^\infty {\hat h}_{l,s}  
\left(\frac{R}{r}\right)^{l+s}, 
\label{h_ext} 
\ee 
and
\be 
{\hat h}_{l,s} =  
\frac{(l+s-2)!(l+s+1)!(2l+1)!}{s!(l-2)!(l+1)!(2l+s+1)!}  
\left(\frac{2M}{R}\right)^s  {\hat h}_{l,0} . 
\ee 
The sum in Eq.~\eqref{h_ext} is the hypergeometric function
$_2F_1\left(l-1,l+2;2 l+2;2M/r\right)$, see \cite{arfken2005}. The factor
${\hat h}_{l,0}$ is arbitrary, as it corresponds to an overall normalization
of the perturbation.  
assuming ${\hat h}_{l,0} =1$.

Matching the interior and exterior solutions for $h_l(r)$ completes the boundary conditions. The first matches the function at the surface, 
\be 
\lim_{\varepsilon\rightarrow 0}  \left[ 
h_l(R-\varepsilon) - h_l(R+\varepsilon) \right] = 0.
\label{cont_cond} 
\ee 
The second, which is given by the condition on the Wronskian, matches the derivatives at the surface,
 \be 
\lim_{\varepsilon\rightarrow 0} \left[ 
h_l(R-\varepsilon) h'_l(R+\varepsilon)  
- h'_l(R-\varepsilon) h_l(R+\varepsilon) 
\right] = 0. 
\label{match_cond} 
\ee 
Both conditions must be true for all values of $l$. 

\section{Numerical Solution}
\label{NumSol}
Following the formulation presented in \cite{Lock03}, as summarized by Sec.~\ref{Formulation} , has effectively changed the problem of finding $\kappa$ from solving the dynamical Einstein equations to solving coupled ODEs for spherical harmonic expansion coefficients. Solving the perturbation Eqs.~\eqref{GR:sph_V}--\eqref{GR:sph_h_0''}, and \eqref{om_th_ph}--\eqref{om_r_th}, with the boundary conditions Eqs.~\eqref{reg_cond}--\eqref{eq:BConW} and \eqref{cont_cond}--\eqref{match_cond} is analytically intractable, but numerically feasible. 

The first step is to solve the OVL equations and find the equilibrium
functions $\lambda(r), \nu(r), \omega(r), p(r)$ and $\epsilon(r)$. Next,
insert the regularity conditions explicitly into the perturbation equations.
Since the eigenfunctions, $h_l, U_l, V_{l+1}, W_{l+1}$, represent coefficients
of an infinite series we must truncate at maximum value for $l$, let us denote
it as $l_{max}$, to get a finite number of equations. Because we are focusing
on axial-led hybrid modes we need to set $l_{max}$ to be a odd number in order to
get a closed system of equations. This choice of axial-led hybrids also means
we solve for the eigenfunctions, $h_l, U_l, W_{l+1}, V_{l+1}$, where $l=m,
m+2 \ldots$, and set the others to zero. Next, note that each term in the
perturbation equations and boundary conditions can be written as the product
of a background (equilibrium) function $B(r)$ and foreground (perturbation)
function $F_l(r)$. For example, in the term  $4 r (\lambda + \nu ) \bar{U}_l(r)$  the background function is $4r (\lambda + \nu ) (r/R)^l$  and the foreground function is $\overline{U_l}$. 

\subsection{Chebyshev-Galerkin Method}
We solve the perturbation equations by expanding both $B(r)$ and $F_l(r)$ in Chebyshev polynomials. Chebyshev polynomials have the form
\be 
T_i(y) = \cos(i\arccos y), \quad i= 1,2,3 , \ldots
\label{chebdef} 
\ee 
and are defined on the domain $[-1,1]$. For our purposes the Chebyshev polynomials' most important property is their exponential convergence when approximating well-behaved functions \cite{NumRec}. In general one can express any well-behaved function $S(y)$ on the domain $[-1,1]$ in terms of these polynomials as
\be
S(y) = \sum_{i=0}^{i_{max}} s_i T_i(y) - \frac{1}{2}s_0, 
\label{genericCheby}
\ee
where $i_{max}$ represents the highest order Chebyshev polynomial that is used to approximate the function.
The coefficients $s_i$ are extracted using the following formula:
\be
s_i = \frac{2}{i_{max}} \sum_{j=0}^{i_{max}+1} S \left[\cos \left(\frac{\pi(j+\frac{1}{2})}{i_{max}} \right) \right] \cos \left( \frac{\pi i (j+\frac{1}{2})}{i_{max}} \right)
\label{ChebyCoef}
\ee
In order to make use of these functions we have to change the domain of our functions from $[0,R]$ to $[-1,1]$ using 
\be 
\label{eq:transform}
y=2\left({r\over R}\right)-1.
\ee 
Thus we transform $B(r)$ into $B(y)$ and expand as,
\be 
B(y) = \sum_{i=0}^{i_{max}} b_i \ T_i(y) - \half b_0, 
\label{gen_back_series} 
\ee 
similarly for $F_l(r)$ we have, 
\be 
F_l(y) = \sum_{i=0}^{i_{max}}  f_{l,i} \ T_i(y) - \half f_{l,0}.
\label{gen_pert_series} 
\ee
Let us define the derivative of $F_l(r)$ as
\be 
F'_l(y) = \frac{d}{dr} F_l =   \sum_{i=0}^{i_{max}}  f'_{l,i} \ T_i(y)  - \half f'_{l,0}.
\label{cheb_deriv_series} 
\ee 
Here the prime notation in $f'_{l,i}$ does not mean derivative, rather it is a numeric coefficient for the derivative expansion of a function.  Similarly we define $F''_l(y) $ for second derivatives of $F_l(r)$, and $f''_{l,i}$ as its coefficients. 

With these definitions in place, the terms in the perturbation equations will be of the form $B(y)F(y)$, or $B(y) F'_l(y)$, or $B(y) F''_l(y)$. There  is a relationship between the $f'_{l_i}$ and $f_{l_i}$ coefficients given by the identity:
\be 
 f'_{l,i} -  f'_{l,i+2} = 4(i+1)f_{l,i+1}. 
\label{cheb_deriv} 
\ee 
It should be noted that Eq.\eqref{cheb_deriv} has a 4 whereas the standard
formula, e.g.\ in \cite{NumRec}, has a 2.
We include the extra factor of 2 to transform $d/dy$ to $d/dr$ via
Eq.\eqref{eq:transform}.
The identity Eq.\eqref{cheb_deriv} can be used twice to find the relationship
between $f''_{l_i}$ and $f_{l_i}$. Since $ f'_{l_i}$ and $f''_{l_i}$ are not
really new coefficients it will be sufficient to say that every term in the
perturbation equations is of the form $B(y) F(y)$. This allows us to use
another Chebyshev identity  \cite{arfken2005},
\be 
B(y) F_l(y) = \half\Biggl[\sum_{i=0}^{i_{max}} \pi_{l,i} \ T_i(y) - \half \pi_{l,0}\Biggr] 
\label{cheb_prod} 
\ee
where 
\be 
\pi_{l,i} = \sum_{j=0}^{i_{max}}  
\biggl[ b_{i+j} + \Theta(j-1) b_{|i-j|} \biggr] f_{l,j}  
\ee 
with 
\be 
\Theta(k) = \left\{ \ba{l} 
0 \ \ \ \ \ \ \ \ \ \ \ \mbox{for $k<0$} \\ 
1 \ \ \ \ \ \ \ \ \ \ \ \mbox{for $k\geq 0$} 
\ea 
\right. ,
\ee
to expand every term in the perturbation equations in Chebyshev polynomials. 

Using the definitions Eqs.~\eqref{gen_back_series}--\eqref{cheb_prod} we can also re-write the boundary conditions in Chebyshev form in the same way

\subsection{Finding $\kappa$}
With all of the terms in the perturbation equations in Chebyshev form we can extract the Chebyshev coefficients using Eq.~\eqref{ChebyCoef}. This will lead to a system of, $2(l_{max} -3) + 4 i_{max}$, algebraic equations for $\kappa$ and coefficients of the unknown functions, $f_{l_i}$. We can schematically represent the system of equations as $A(\kappa) x = 0$, where $A$ is a matrix and $x$ is the vector
\begin{widetext}
\be
x = \left [ \overline h_{l_0} \hspace{2mm} \overline h_{l_{1}}  \ldots \overline h_{l_{i_{max}}}  \ldots  \overline U_{l_0}  \hspace{2mm}  \overline U_{l_{1}} \ldots \overline U_{l_{i_{max}}}  \ldots  \overline W_{{(l+1)}_0} \hspace{2mm} \overline W_{{(l+1)}_{1}} \ldots \overline W_{{(l+1)}_{i_{max}}}    \ldots  \overline V_{{(l+1)}_0} \hspace{2mm}  \overline V_{{(l+1)}_{1}} \ldots \overline V_{{(l+1)}_{i_{max}}}  \right ] .
\ee
\end{widetext}
Before finding $\kappa$ we must incorporate the $\frac{3}{2}(l_{max} - 2) +3$ equations that come from converting the boundary conditions into Chebyshev form. To do this we replace the equation that came from the highest order extracted coefficient $\pi_{imax}$ for each eigenfunction with a boundary condition.  

We solve for $\kappa$ using the condition $\det(A(\kappa)) =0$. This leads to
a high degree ($\mathcal{O}(500+)$) polynomial  equation for $\kappa$. Finding the roots for such a polynomial is difficult when using standard root-finders. Therefore we created two root-finding algorithms which incorporated and went beyond some of the standard root-finding techniques. The key idea of one, is to solve for the roots of the function: $\tan^{-1}\log | \det(A(\kappa) ) |$ instead. 
The key idea of the other is use the decomposition, $SVD(A) = \{{\cal{U}}, \Sigma, {\cal{V}}\} $,  and find the value of $\kappa$ that results in the smallest value for last element on the diagonal of $\Sigma$. 

\begin{figure}[ht]
\begin{center}
\includegraphics[scale=0.95]{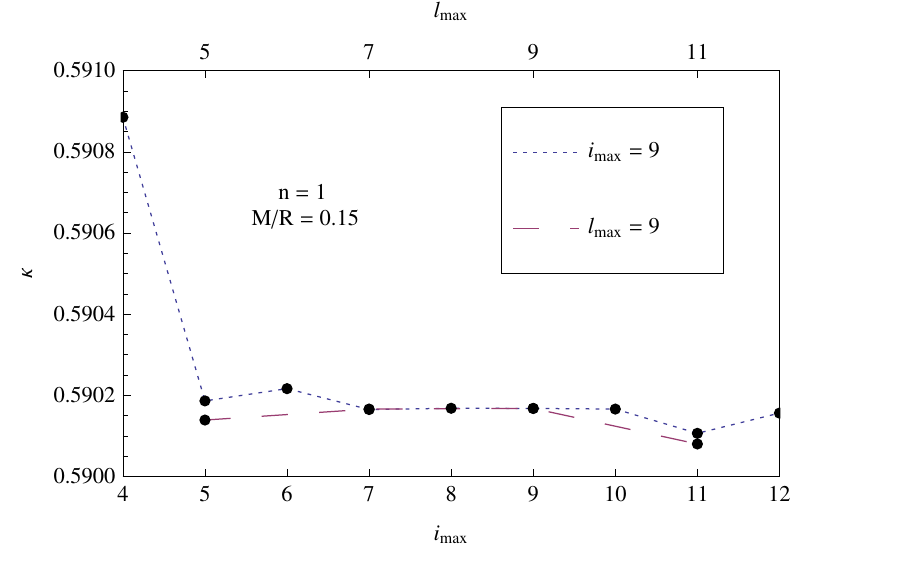}
\end{center}
\caption{Convergence of eigenvalue for $n=1$ polytropic EoS. The short (blue) dashes show convergence in $i_{max}$ while fixing $l_{max} = 9$, and the longer (purple) dashes show the convergence in $l_{max}$ when fixing $i_{max} =9$. The divergence of the eigenvalue sets for higher values of $l_{max}$ and $i_{max}$ due to finite precision, see text.}
\label{fig:imaxCnvrg}
\end{figure}

Both root finders achieved convergence at the fourth decimal place for polytropic EoS, and the third decimal place for realistic EoS. The loss of precision comes form the fact that realistic EoS have to be numerically interpolated, whereas polytropic EoS have analytical forms, see Sec.~\ref{Conclusions}. The first algorithm mentioned was used for the results that appear below, because it was easier to automate. In Fig.~\ref{fig:imaxCnvrg}  we show the convergence for an $n=1$ polytrope with compactness of $.150$. Although this figure does show that our code quickly converges, it obscures the fact that in practice one should increase $i_{max}$ and $l_{max}$ in step to get convergence.  This figure also shows that the eigenvalue stops converging above $l_{max} =11$  and $i_{max} =9$. 
For tabulated EoS, a minimum of $l_{max} = 15$  and $i_{max} =13$ are required for convergence. Divergence sets in for tabulated EoS at $l_{max} =21$  and $i_{max} =19$. The convergence stops after certain values due to truncation errors and the high orders of the polynomials. 

The issue with both root-finders is that they lead to multiple roots. The way to determine which root is correct is to start $l_{max}$ and $i_{max}$ at small values so that one only finds $3$ to $5$ roots, and then keep the root(s) closest to the Newtonian estimate of $\kappa_0 = 2/3$. Then as $l_{max}$ and $i_{max}$ are increased the correct root will converge whereas the others will change unpredictably. 

\subsection{The eigenfunctions}
Our procedure for finding the eigenfunctions is as follows. First rewrite all of the variables in the set of equations, $A(\kappa) x = 0$, in terms of one variable, e.g. $h_{2_0}$. Then impose a normalization condition such that $h_{2_0}=1 $ at the surface of the star. With this solution for $h_{2_0}$ at hand, populate the rest of  $\vec{x}$. 

There is an issue with solution for the eigenfunctions, $h_l, U_l, W_l, V_l$
for stars near the maximum compactness, i.e.\ the maximum mass of a star
stable against radial collapse.
Very close to the maximum mass, the solutions develop extra maxima and minima.
Evidently the highest-order Chebyshev coefficients are acquiring spuriously
high values.
We conjecture that this is due to the sensitivity of the stellar equilibrium
functions to small perturbations near the maximum mass.
This does not seem to significantly affect the eigenvalues though, which
continue their smooth trend as functions of compactness.

%
\section{Results}
\label{Results}
\subsection{The n = 1 polytrope}
We have successfully reproduced results presented in \cite{Lock03} for an
$n=1$ polytrope. They found that for a compactness of $M/R=.15$, $\kappa =
.5901$. We calculated $\kappa = .5902$ for such a star, which represents a
difference of only $.01\%$. This small discrepancy could have come from
the precision of the code that was used in the calculations (we used {\sc
Mathematica} standard precision, which is somewhat better than double
precision). Also that is about the current level of uncertainty in the CODATA
standard value of $G$~\cite{Gnow} and the level of changes in that value in
recent years~\cite{Gthen}.
The eigenfunctions we calculated for the $\kappa = .5902$ \emph{r}-mode are
exactly the same as those present in Fig.~3 and 4 of \cite{Lock03}, as far as
the eye can see.  

\begin{figure}[htbp]
\begin{center}
\includegraphics[width=.5\textwidth]{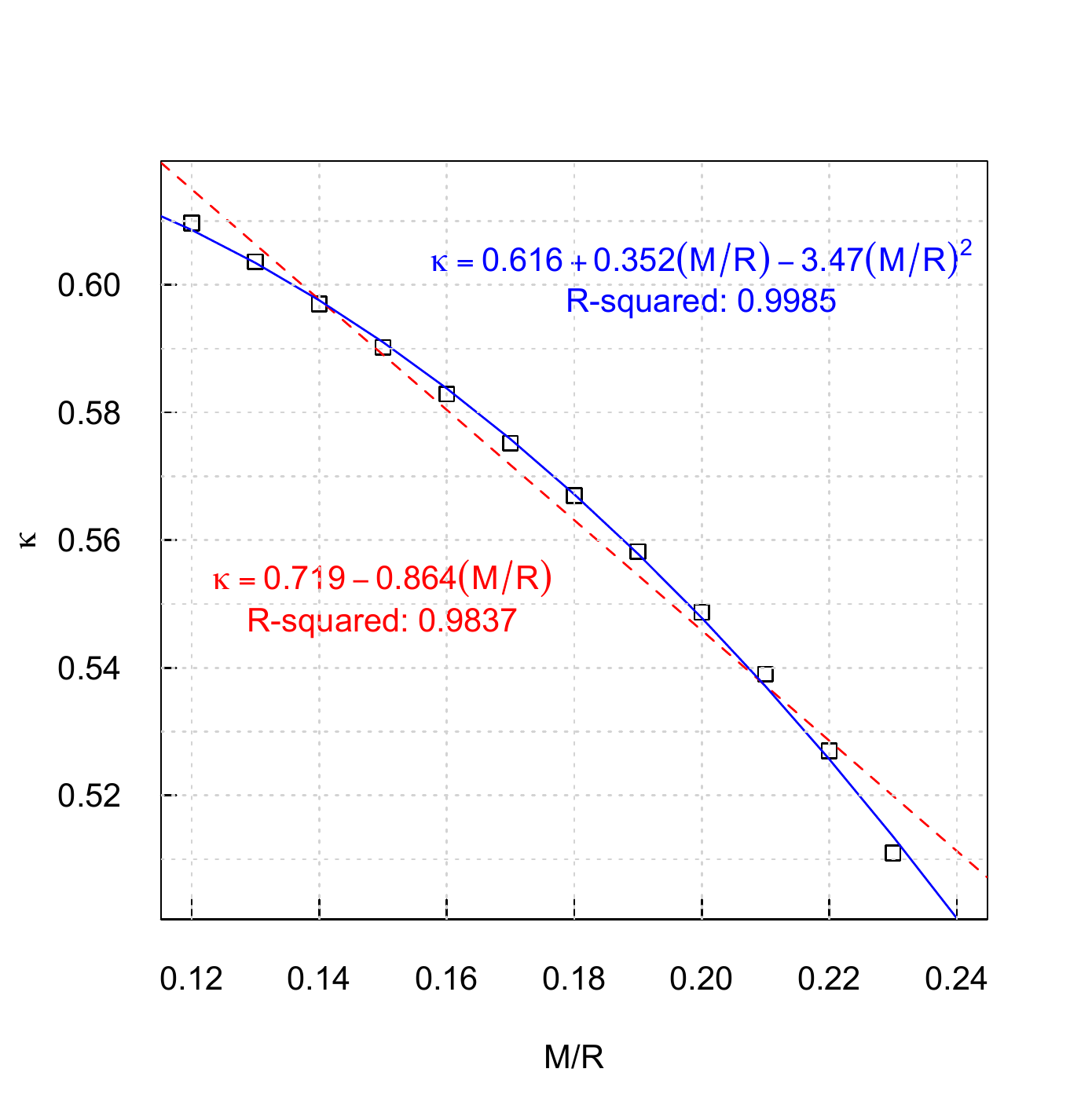}
\end{center}
\caption{The eigenvalue, $\kappa$, for an equilibrium stellar mass sequence of
an n=1 polytrope. The radius was kept fixed at 12.53 km, and the mass changed
ranged from 1.01--1.95 $M_\odot$. Linear and quadratic fits to the data are also presented.}
\label{fig:Rplotn=1data}
\end{figure}

We have extended the results of \cite{Lock03} for an $n=1$ polytrope by examining how the \emph{r}-mode frequency changes as compactness of the star is varied. The results are shown in Fig.~\ref{fig:Rplotn=1data}.  The compactness was changed by increasing the mass, from 1.01--1.95 $M_{\odot}$, while holding the radius fixed at 12.53 km. 
It is clear from the plot that $\kappa$ decreases as compactness increases. In
the plot we include least-squares fits to a linear and quadratic model. The
$R^2$ values indicate that the quadratic model
\begin{equation}
\kappa = 0.616 + 0.352 (M/R) - 3.47 (M/R)^2
\end{equation}
is a better fit for the data. 

It seems that the negative coefficient for the quadratic term is a generic result. The same result was found by \cite{Lock03} for the stellar equilibrium sequence of an $n=0$ polytrope. The results for polytropic EoS are a useful guide when examining the results from the realistic or tabulated EoS, because tabulated EoS do not have analytical form like the polytropes but rather one has to use the interpolation methods mentioned in Sec.~\ref{Formulation}.

\subsection{Tabulated Equations of State}
\begin{figure*}[htbp]
\begin{center}
\includegraphics[scale=0.85]{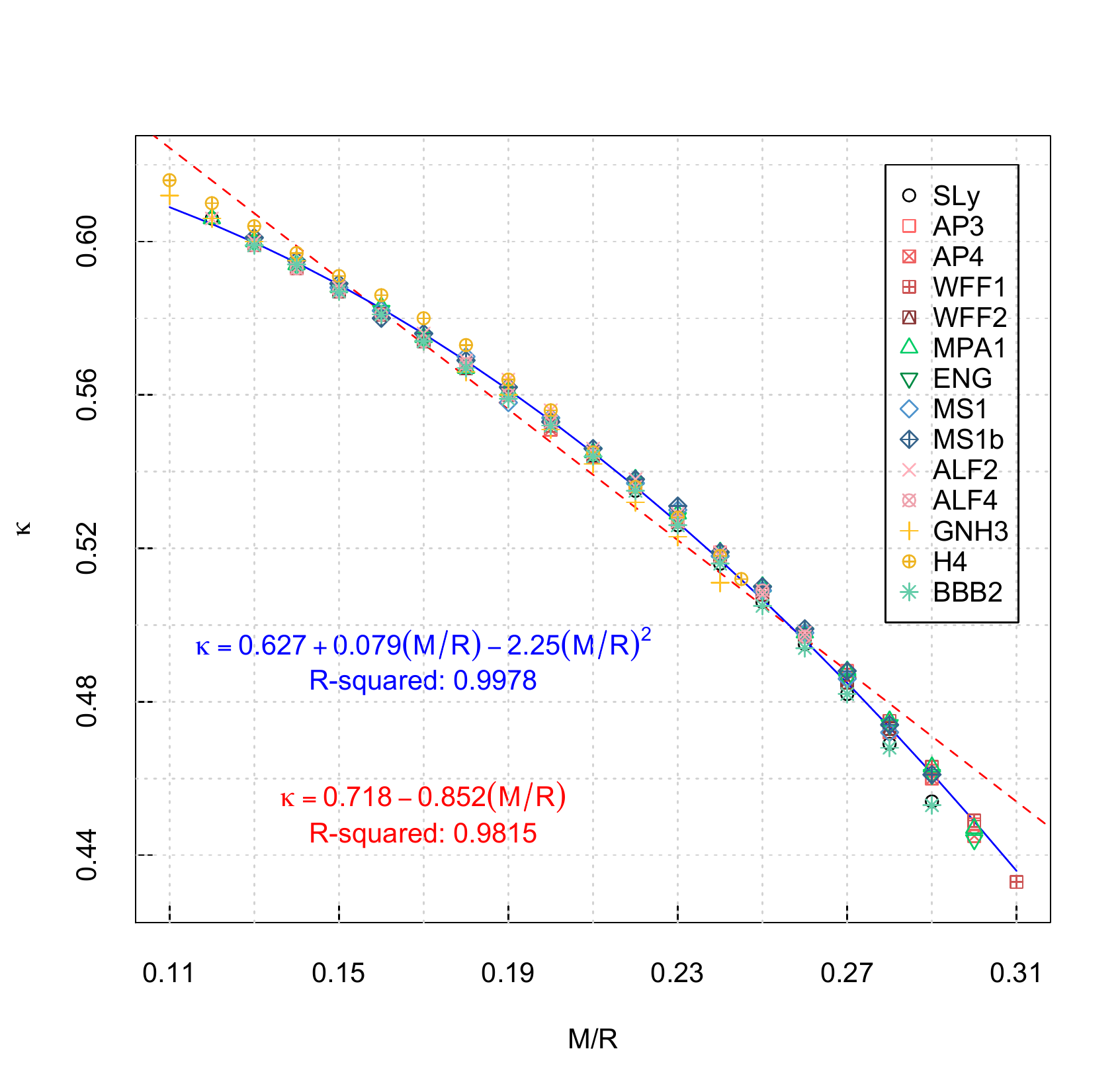}
\caption{The values of $\kappa$ for all 14 EoS under consideration. The dashed (red) line represents linear fit, and the solid (blue) line represents the quadratic fit. Ordinary-least-squares regression was used to get both the linear and quadratic fit. The equations for the linear and quadratic fit along with their $R^2$ for the fits are also presented.}
\label{fig:RplotKappa}
\end{center}
\end{figure*}

Figure \ref{fig:RplotKappa} contains the values of $\kappa$ for all 14 EoS under consideration. These were chosen from a standard list of EoS used by \cite{LindIndik}, and \cite{Read4P}, under the constraint that the EoS could support a 1.85 $M_{\odot}$ star, see Sec.~\ref{sec:intro}. The lowest mass that was used for any EoS was 1.02 $M_{\odot}$, the maximum mass used in any calculation was 2.76 $M_\odot$.%

There are three main things to notice in Fig.~\ref{fig:RplotKappa}. One, the
values of $\kappa$ decrease as the compactness of the star increases. Two, the
generic shape of the data is parabolic, as shown by the solid (blue) fitted
curve.  Ordinary-least-squares regression was used to get both the linear and
quadratic fit.  Examining the $R^2$ value we see again that the quadratic
model
\begin{equation}
\kappa = 0.627 + 0.079 (M/R) - 2.25 (M/R)^2
\end{equation}
is a better fit to the data. We also calculated the root mean square
error (RMSE) \cite{kutner2005applied}, as a way to quantify the deviation of
the individual EoS from the quadratic fit. These values are presented in Table
I. The total RMSE for all data points is $2.02 \times10^{-3}$. Three, the
range of $\kappa$ for tabulated EoS is larger than that for $n=1$ polytropic
model. This is mainly due to the fact that realistic equations are less stiff
than the $n=1$ polytrope, and so one can squeeze more mass into the same
radius, thus increasing the maximum compactness from $.220$ to $.310$ for some
\emph{realistic} EoS. This greater range for $\kappa$ has importance
implications for the applications of our results, and these will be explored
further in the next section.

\begin{figure*}
        \centering
        \begin{subfigure}[b]{0.5\textwidth}
                \includegraphics[width=\textwidth]{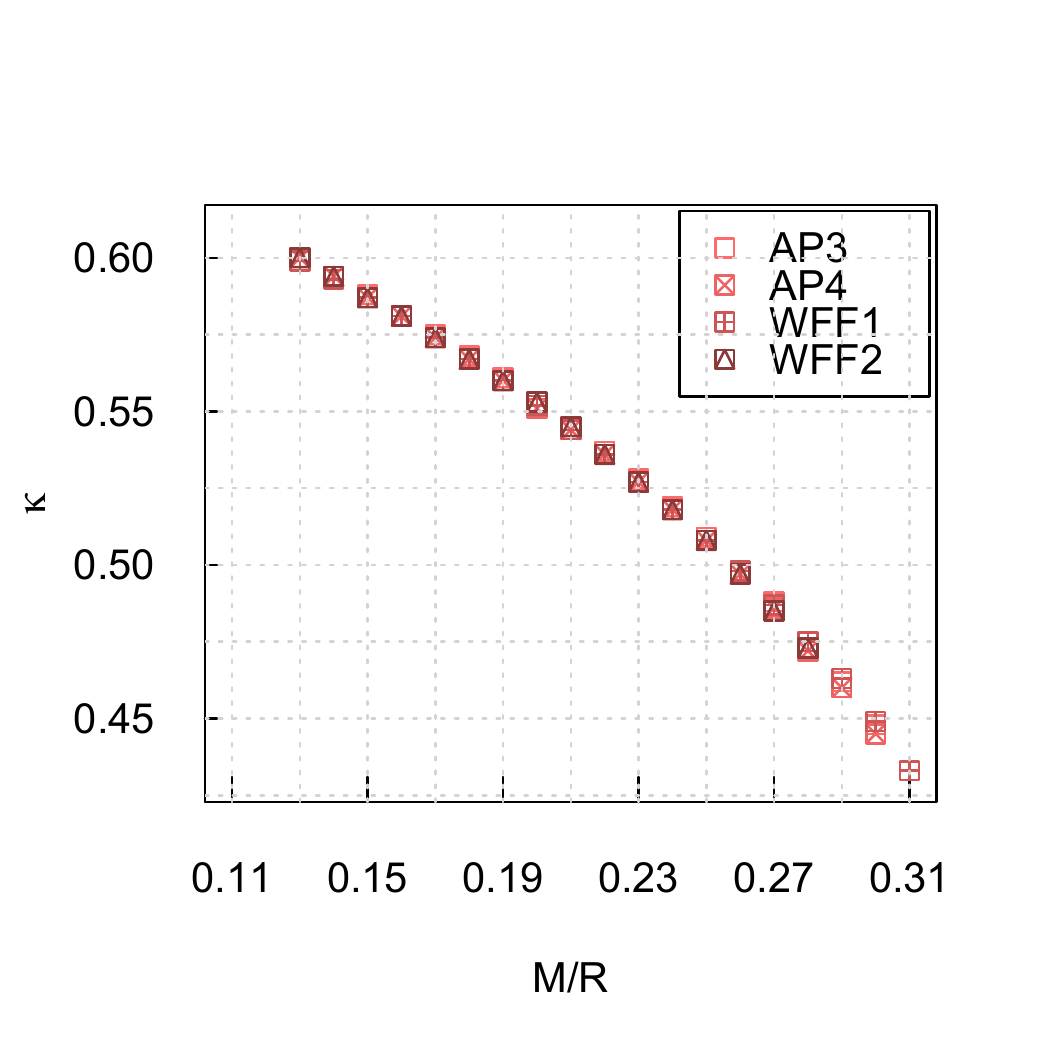}
                \caption{}
                \label{fig:RplotVarEOS}
        \end{subfigure}%
        ~ 
        \begin{subfigure}[b]{0.5\textwidth}
                \includegraphics[width=\textwidth]{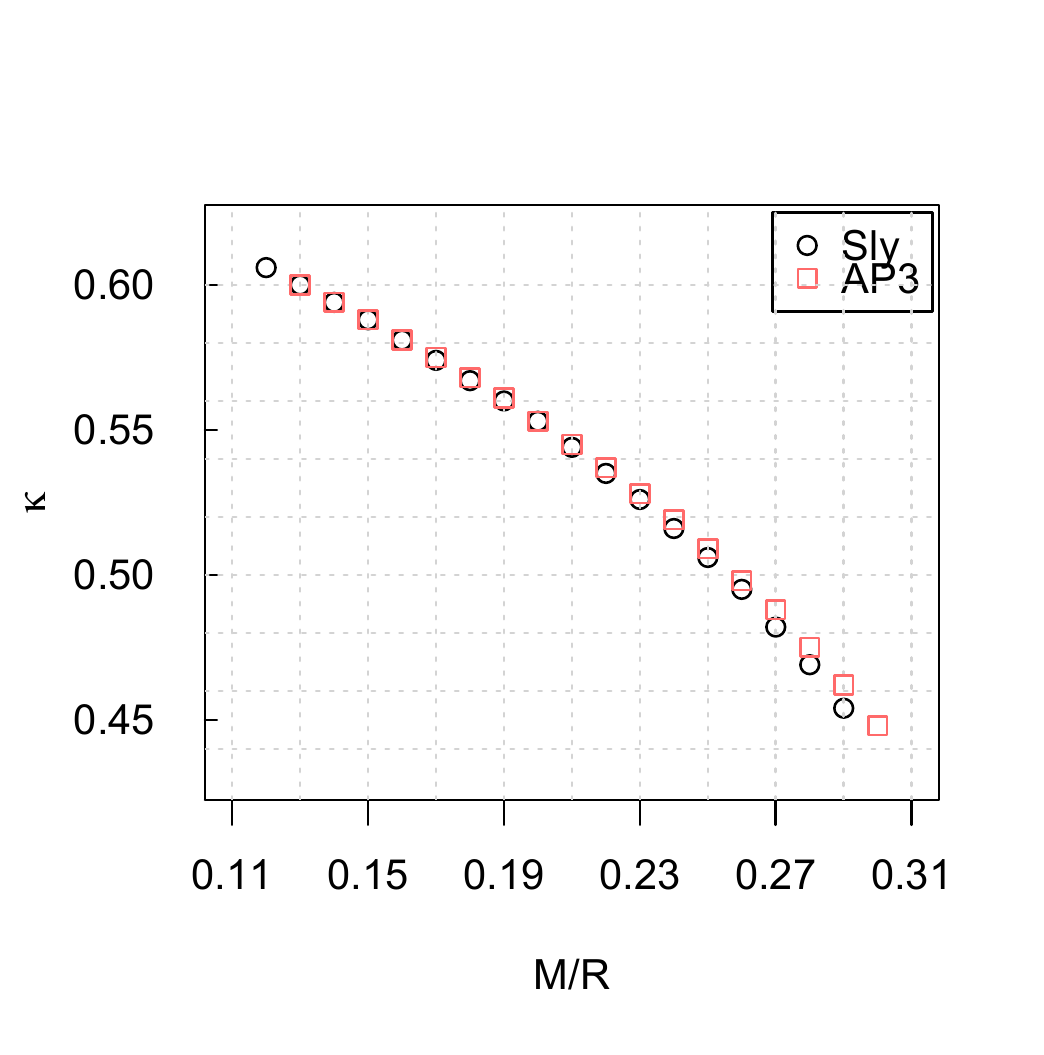}
                \caption{}
                \label{fig:RplotSLYandAP3}
        \end{subfigure}%
        \caption{In (a) we compare the values of $\kappa$  for EoS derived using variational methods. In (b) we compare the values of $\kappa$ for SLy and AP3 EoS.}
        \label{fig:EOSFamilies}
\end{figure*}

To further examine the results for the tabulated EoS we have split
Fig.~\ref{fig:RplotKappa} into two plots.
Figure~\ref{fig:EOSFamilies}(a) shows a plot for a family of variational
method EoS, and shows that the value of $\kappa$ does not change very much
within this family.
Figure \ref{fig:EOSFamilies}(b) shows the difference between a member of this
family and an EoS derived by very different methods.
It examines whether our results can be used to
constrain or rule out certain EoS. The plots show how the range of $\kappa$
for SLy \cite{Douchin} (which is the only EoS in its family) relates to that
for AP3 \cite{APR} from the family shown in Fig.~\ref{fig:EOSFamilies}(a).
These graphs will be discussed further in the next section. 

Finally we present two tables. Table \ref{tab:shortData} lists all the tabulated EoS that were
used, along with the stable maximum mass of a non-rotating star, the radius of a
$1.4 M_{\odot}$ star in the sequence, $\kappa$ for a compactness of $.15$, the
coefficients for the quadratic fit of the $\kappa$ values for that EoS, and
the $RMSE$. The $R^2$ value for each fit (not shown in the table) ranged from .9986 to .9998, again
showing that the quadratic model is a good fit for the data.  From this fit we
see that the quadratic term can be up to a few percent of the frequency. This
has significant implications for GW searches, as well as attempts to measure compactness from an \emph{r}-mode frequency. Table \ref{tab:longData} gives the numerical values plotted in Fig.~\ref{fig:RplotKappa}. 
\begin{table}[!h]
\begin{center}
\begin{tabular}{l | c c | c | c c c | c}
\hline\hline
EoS  & $M_{\rm max}$ & $R_{1.4}$ & $\kappa_{.15}$ & a  & b &  c & RMSE \\
&  &  & &  &  &  & $1 \times 10^{-3}$  \\
\hline
Sly& 2.049& 11.736 &0.587 & 0.622  & 0.151 &-2.48 & 2.35  \\
AP3& 2.390&12.094 & 0.588  & 0.619 &  0.142  & -2.36 & 1.39 \\
AP4& 2.213& 11.428 & 0.587& 0.626 & 0.150 & -2.41 & 1.56 \\ 
WFF1& 2.133& 10.414 &0.587 & 0.617 & 0.160 & -2.40 & 1.54 \\ 
WFF2& 2.198&11.159 & 0.587& 0.628 & 0.060 &  -2.17  & 1.08 \\
MPA1& 2.461& 12.473 & 0.588& 0.629 & 0.052 & -2.14 & 1.70 \\
ENG& 2.240& 12.059 & 0.588 & 0.613 & 0.215 & -2.56 & 1.51 \\
MS1& 2.767& 14.918 & 0.588 & 0.624 & 0.107 &  -2.30  & 1.62 \\
MS1b& 2.776& 14.583 & 0.589  & 0.619 & 0.156 & -2.40 & 1.98 \\
ALF2& 2.086& 13.188 & 0.588& 0.632  &0.026 & -2.08 & 1.63 \\
ALF4& 1.943& 11.667 & 0.587 & 0.632 &   0.002 & -1.98 & 1.25 \\
GNH3&1.962& 14.203 & 0.587& 0.639  &   -0.027 & -2.09 & 2.80 \\
H4& 2.032& 13.774 &0.591 & 0.640 &- 0.002  & -2.13 & 3.65 \\
BBB2&1.918&11.139 & 0.587,& 0.611 &  0.249 &  -2.71 & 2.83 \\
\hline
\end{tabular}
\end{center}
\caption{We present a list of all the tabulated EoS, for each EoS we show the
stable maximum mass, the radius for a $1.4 M_\odot$ star, $\kappa$ for a compactness of $.15$, the coefficients for the quadratic fit of the $\kappa$ of the form $a + b\left( \frac{M}{R} \right)  +c \left(\frac{M}{R}\right)^2 $, and the root mean square error (RMSE) of the EoS data points to the quadratic fit. }
\label{tab:shortData}
\end{table}

\section{Conclusions}
\label{Conclusions}
We begin with a discussion of the $n=1$ polytrope results. The fact that the \emph{r}-mode frequencies go down as compactness increases might come as a surprise. However, one must remember to keep track of reference frames. In the reference frame of the star $\kappa$ does decrease because the restoring force is proportional to $\bar\omega(r) = \Omega - \omega$, and this goes down as compactness is increased. But the observed frequency in the inertial frame is $|\kappa - m |\Omega$, see Eq.~\eqref{eq:kappa_def}, and this increases as $\kappa$ decreases. 

To compare the results from the $n=1$ polytrope, and tabulated EoS we can
examine Fig.~\ref{fig:Rplotn=1data} and \ref{fig:RplotKappa}. From this we
notice that the $\kappa$ for tabulated EoS can differ from those of the
polytropic model by of order ten percent in frequency. This shows the need to use realistic EoS when calculating \emph{r}-mode frequencies.

Let us now discuss the implications of the parabolic shape of the $\kappa$ values shown in Fig.~\ref{fig:Rplotn=1data}, and \ref{fig:RplotKappa}. Overlaid on these plots are two fits from which it is clear that the quadratic fit is better than the linear one. These figures show that the corrections to the \emph{r}-mode frequency using post-Newtonian approximations must be carried out to at least second order. For example, the first order post-Newtonian formula for an $n=0$ polytrope given by \citet{Lock03} is 
\be
\kappa_{pN} = \frac{2}{m+1} \left [ 1 - \frac{ 8 (m-1) (2m +11)}{5 (2m+1) (2m +5) } \frac{M}{R} \right ] ,
\ee
for $m=2$ this gives
\be
\kappa_{pN} = \frac{2}{3} \left(1-\frac{8}{15} \frac{M}{R}\right).
\ee
Figure \ref{fig:Rplotn=1data} and \ref{fig:RplotKappa} show that the equation above is insufficient since it does not account for quadratic, $(M/R)^2$, corrections. 

Moving on to the astrophysical applications of our results. Gathering
data from \emph{realistic} EoS we see that the range of $\kappa
$ is approximately 0.614--0.433 for compactness values 0.110--0.310. If the spin
frequency of a star is known from electromagnetic observations, this range can
be used to conduct narrow-band gravitational wave searches for known pulsars.
Using Eq.~\eqref{eq:kappa_def} we see that our range of $\kappa$ gives the
range $1.39 \Omega < \sigma_I < 1.57 \Omega$, where $\sigma_I$ is the
frequency observed in our reference frame. Taking the Crab pulsar
($\Omega/2\pi = 29.7$ Hz, \cite{ATNF}) as an example, our range of $\sigma_I$ suggests a
narrow-band search for \emph{r}-modes be carried out from 41.3--46.6~Hz.

Alternatively, our results will be of use in the case where a GW detection is
made for an electromagnetically unknown pulsar.  Suppose the gravitational
wave signal has a frequency of  100 Hz. Astronomers can search for the pulses
at 50 Hz, assuming the signal came from a non-axisymmetric deformation in the
star, and in the range of 63.7--71.9~Hz, assuming the signal came from
\emph{r}-modes of the star. 

Another potential use for this research that was mentioned in the introduction
was the ability use the \emph{r}-mode frequencies to constrain the nuclear
equation of state. We see from Fig.~\ref{fig:EOSFamilies}(a) that
\emph{r}-mode detections alone will not be enough to distinguish between
members of an EoS family.  Figure  \ref{fig:EOSFamilies}(b) shows a slightly
more promising result. This figure shows that it may be possible to
distinguish between different EoS families. However, this would be made
difficult without additional electromagnetic data on quantities such as the compactness. This is due to the scale of the deviations seen in Fig.~\ref{fig:EOSFamilies}(b), which are less than 1\%. These deviations to $\kappa$ have to compete with the physical phenomena described in Sec.~\ref{sect:effects} which for most stars can also change  $\kappa$ by 1\%.  Therefore it will be impossible to distinguish what is truly giving rise to the change in $\kappa$ without more research into these effects. 

In this context it is interesting to note the  recent  report in
\cite{2013r-modeDiscovery} of the possible detection of an \emph{r}-mode in
the outburst of the accreting millisecond pulsar XTE J1751-305.  Reference
\cite{Andersson2014} showed that the observed frequency of the oscillation, if
interpreted as the \emph{r}-mode frequency, gave rise to a sensible constraint
on the mass-radius relation for the star, see their Fig.~1.  Reference
\cite{Andersson2014} included both relativistic corrections and rotational
ones in their analysis, but the former were based on the uniform-density
calculations of \cite{Lock03}, rather than realistic EoS of the sort
considered in this paper.  However, comparison of the plot of $\kappa$ verses
compactness  in Fig.~1 of  \cite{Andersson2014}  with Fig.~\ref{fig:RplotKappa} above shows
that the analysis  would not change significantly if it were repeated using
realistic EoS, something to be expected given the rather narrow variation of
$\kappa$ with EoS shown in our Fig.~3.
The situation is broadly similar for the second possible neutron-star $r$-mode
detection \cite{r-modeAnotherDiscovery}.

To sum up, we have been successful in finding a range of \emph{r}-mode
frequencies for both polytropic and tabulated EoS. Furthermore we have shown
that our results can be used as input data for electromagnetic and GW searches.  Along with these successes there are some issues that should be discussed.  

One issue we encountered is that the precision of the $\kappa$ values decreased by an order of magnitude for the tabulated EoS compared to polytropes. We believe this comes from the fact that the tabulated EoS have to be interpolated instead of coming in analytical form like polytropes. From our discussion in Sec.~\ref{sec:equilSol} we know that the different interpolation schemes results in percent differences 0.1\%. This shows that when using tabulated values we can only hope to get up to three significant digits, regardless of the interpolation scheme used. It is difficult to address this issue since it is an inherent problem with tabulated EoS. Perhaps in future work we can use analytical fits to the tabulated EOS such as those presented in \cite{Read4P} and \cite{LindIndik} to mitigate this issue. 

Another issue is of course the various physical phenomena that were ignored. These will inevitably have some impact upon the \emph{r}-mode frequency, even if our simple estimates indicated that relativistic effects were likely to be the most important. Therefore it is still necessary to explore the effects of other mechanisms.  In particular, in the future we would like verify that the corrections that come from rapid rotation are significant only for very rapidly rotating stars, and that the crustal effects are important only in narrow spin frequency bands. Including these effects would allow us to distinguish between families of EoS if we get compactness information.

\section{Acknowledgements}
The authors would like to thank Ruxandra Bondarescu, Andrew Lundgren, and John Friedman for their helpful conversations about the
material in this paper. We would also like to thank Ra Inta for considerable
helpful feedback on the manuscript, and Matthew Pitkin and Michal Bejger for
careful readings of the penultimate draft. AI and BJO acknowledge support from NSF
grant number PHY-1206027. DIJ acknowledges support from STFC via grant number
ST/H002359/1. DIJ and BJO acknowledge travel support from CompStar (a
COST-funded Research Networking Programme). This paper has document number
LIGO-P1400218.

\begin{table*}[h]
\begin{center}
\begin{tabular}{c | c c c c c c c c c c c c c c}
\hline
\hline
$M/R$ &SLY&AP3&AP4&WFF1&WFF2&MPA1&ENG&MS1&MS1B&ALF2&ALF4&GNH3&H4&BBB2\\
\hline
0.11&-&-&-&-&-&-&-&-&-&-&-&0.612&0.616&- \\
0.12&0.606&-&-&-&-&0.606&-&-&-&0.606&-&0.606&0.610&- \\
0.13&0.600&0.600&0.600&0.599&0.600&0.600&0.600&0.601&0.601&0.600&0.599&0.600&0.604&0.599 \\
0.14&0.594&0.594&0.594&0.593&0.594&0.594&0.594&0.595&0.595&0.595&0.593&0.595&0.597&0.594 \\
0.15&0.588&0.588&0.587&0.587&0.587&0.588&0.588&0.588&0.589&0.588&0.587&0.587&0.591&0.587 \\
0.16&0.581&0.581&0.581&0.581&0.581&0.583&0.582&0.582&0.58&0.581&0.581&0.581&0.586&0.581\\
0.17&0.574&0.575&0.574&0.574&0.574&0.575&0.575&0.576&0.576&0.576&0.574&0.574&0.58&0.574 \\
0.18&0.567&0.568&0.567&0.567&0.567&0.567&0.568&0.57&0.569&0.569&0.568&0.566&0.573&0.567 \\
0.19&0.56&0.561&0.56&0.56&0.56&0.561&0.561&0.558&0.562&0.564&0.56&0.56&0.564&0.559 \\
0.20&0.553&0.553&0.551&0.552&0.553&0.554&0.553&0.554&0.553&0.556&0.553&0.551&0.556&0.552 \\
0.21&0.544&0.545&0.544&0.544&0.545&0.545&0.545&0.546&0.546&0.546&0.545&0.542&0.545&0.544 \\
0.22&0.535&0.537&0.536&0.536&0.536&0.538&0.537&0.537&0.538&0.538&0.536&0.532&0.536&0.535 \\
0.23&0.526&0.528&0.527&0.527&0.527&0.529&0.528&0.53&0.531&0.529&0.527&0.523&0.528&0.526 \\
0.24&0.516&0.519&0.518&0.518&0.518&0.519&0.518&0.518&0.519&0.519&0.518&0.511&0.518&0.516 \\
0.25&0.506&0.509&0.508&0.508&0.508&0.51&0.509&0.509&0.51&0.508&0.509&-&0.512&0.505 \\
0.26&0.495&0.498&0.497&0.498&0.497&0.498&0.498&0.498&0.499&0.498&0.497&-&-&0.494 \\
0.27&0.482&0.488&0.486&0.487&0.485&0.488&0.486&0.486&0.488&-&-&-&-&0.482 \\
0.28&0.469&0.475&0.472&0.475&0.473&0.475&0.474&0.472&0.474&-&-&-&-&0.468 \\
0.29&0.454&0.462&0.46&0.463&-&0.463&0.461&-&0.461&-&-&-&-&0.453 \\
0.30-&-&0.448&0.445&0.449&-&0.447&0.444&-&-&-&-&-&-&- \\
0.31-&-&-&-&0.433&-&-&-&-&-&-&-&-&-&- \\
\hline
\end{tabular}
\end{center}
\caption{This table gives the numerical $\kappa$ for all tabulated EoS, over the range of compactness values considered in our analysis. These values are plotted in in Fig.~\ref{fig:RplotKappa}. The ``-" indicates a compactness that could not be obtained with that EoS.}
\label{tab:longData}
\end{table*}%

\bibliographystyle{apsrev4-1}
\bibliography{EIGENrefs}	

\begin{thebibliography}{49}%
\makeatletter
\providecommand \@ifxundefined [1]{%
 \@ifx{#1\undefined}
}%
\providecommand \@ifnum [1]{%
 \ifnum #1\expandafter \@firstoftwo
 \else \expandafter \@secondoftwo
 \fi
}%
\providecommand \@ifx [1]{%
 \ifx #1\expandafter \@firstoftwo
 \else \expandafter \@secondoftwo
 \fi
}%
\providecommand \natexlab [1]{#1}%
\providecommand \enquote  [1]{``#1''}%
\providecommand \bibnamefont  [1]{#1}%
\providecommand \bibfnamefont [1]{#1}%
\providecommand \citenamefont [1]{#1}%
\providecommand \href@noop [0]{\@secondoftwo}%
\providecommand \href [0]{\begingroup \@sanitize@url \@href}%
\providecommand \@href[1]{\@@startlink{#1}\@@href}%
\providecommand \@@href[1]{\endgroup#1\@@endlink}%
\providecommand \@sanitize@url [0]{\catcode `\\12\catcode `\$12\catcode
  `\&12\catcode `\#12\catcode `\^12\catcode `\_12\catcode `\%12\relax}%
\providecommand \@@startlink[1]{}%
\providecommand \@@endlink[0]{}%
\providecommand \url  [0]{\begingroup\@sanitize@url \@url }%
\providecommand \@url [1]{\endgroup\@href {#1}{\urlprefix }}%
\providecommand \urlprefix  [0]{URL }%
\providecommand \Eprint [0]{\href }%
\providecommand \doibase [0]{http://dx.doi.org/}%
\providecommand \selectlanguage [0]{\@gobble}%
\providecommand \bibinfo  [0]{\@secondoftwo}%
\providecommand \bibfield  [0]{\@secondoftwo}%
\providecommand \translation [1]{[#1]}%
\providecommand \BibitemOpen [0]{}%
\providecommand \bibitemStop [0]{}%
\providecommand \bibitemNoStop [0]{.\EOS\space}%
\providecommand \EOS [0]{\spacefactor3000\relax}%
\providecommand \BibitemShut  [1]{\csname bibitem#1\endcsname}%
\let\auto@bib@innerbib\@empty
\bibitem [{\citenamefont {{Papaloizou}}\ and\ \citenamefont
  {{Pringle}}(1978)}]{1978PP}%
  \BibitemOpen
  \bibfield  {author} {\bibinfo {author} {\bibfnamefont {J.}~\bibnamefont
  {{Papaloizou}}}\ and\ \bibinfo {author} {\bibfnamefont {J.~E.}\ \bibnamefont
  {{Pringle}}},\ }\href@noop {} {\bibfield  {journal} {\bibinfo  {journal}
  {Mon. Not. R. Astron. Soc.}\ }\textbf {\bibinfo {volume} {182}},\ \bibinfo
  {pages} {423} (\bibinfo {year} {1978})}\BibitemShut {NoStop}%
\bibitem [{\citenamefont {Chandrasekhar}(1970)}]{Chandra}%
  \BibitemOpen
  \bibfield  {author} {\bibinfo {author} {\bibfnamefont {S.}~\bibnamefont
  {Chandrasekhar}},\ }\href {\doibase 10.1103/PhysRevLett.24.611} {\bibfield
  {journal} {\bibinfo  {journal} {Phys. Rev. Lett.}\ }\textbf {\bibinfo
  {volume} {24}},\ \bibinfo {pages} {611} (\bibinfo {year} {1970})}\BibitemShut
  {NoStop}%
\bibitem [{\citenamefont {{Friedman}}\ and\ \citenamefont
  {{Schutz}}(1978)}]{FriedmanSchutz78}%
  \BibitemOpen
  \bibfield  {author} {\bibinfo {author} {\bibfnamefont {J.~L.}\ \bibnamefont
  {{Friedman}}}\ and\ \bibinfo {author} {\bibfnamefont {B.~F.}\ \bibnamefont
  {{Schutz}}},\ }\href {\doibase 10.1086/156143} {\bibfield  {journal}
  {\bibinfo  {journal} {\apj}\ }\textbf {\bibinfo {volume} {222}},\ \bibinfo
  {pages} {281} (\bibinfo {year} {1978})}\BibitemShut {NoStop}%
\bibitem [{\citenamefont {{Andersson}}(1998)}]{Ander}%
  \BibitemOpen
  \bibfield  {author} {\bibinfo {author} {\bibfnamefont {N.}~\bibnamefont
  {{Andersson}}},\ }\href {\doibase 10.1086/305919} {\bibfield  {journal}
  {\bibinfo  {journal} {\apj}\ }\textbf {\bibinfo {volume} {502}},\ \bibinfo
  {pages} {708} (\bibinfo {year} {1998})}\BibitemShut {NoStop}%
\bibitem [{\citenamefont {Friedman}\ and\ \citenamefont
  {Morsink}(1998)}]{FMr-mode}%
  \BibitemOpen
  \bibfield  {author} {\bibinfo {author} {\bibfnamefont {J.~L.}\ \bibnamefont
  {Friedman}}\ and\ \bibinfo {author} {\bibfnamefont {S.~M.}\ \bibnamefont
  {Morsink}},\ }\href {http://stacks.iop.org/0004-637X/502/i=2/a=714}
  {\bibfield  {journal} {\bibinfo  {journal} {\apj}\ }\textbf {\bibinfo
  {volume} {502}},\ \bibinfo {pages} {714} (\bibinfo {year}
  {1998})}\BibitemShut {NoStop}%
\bibitem [{\citenamefont {{Lindblom}}\ \emph {et~al.}(1998)\citenamefont
  {{Lindblom}}, \citenamefont {{Owen}},\ and\ \citenamefont
  {{Morsink}}}]{LindOwenMors98}%
  \BibitemOpen
  \bibfield  {author} {\bibinfo {author} {\bibfnamefont {L.}~\bibnamefont
  {{Lindblom}}}, \bibinfo {author} {\bibfnamefont {B.~J.}\ \bibnamefont
  {{Owen}}}, \ and\ \bibinfo {author} {\bibfnamefont {S.~M.}\ \bibnamefont
  {{Morsink}}},\ }\href {\doibase 10.1103/PhysRevLett.80.4843} {\bibfield
  {journal} {\bibinfo  {journal} {Phys. Rev. Lett.}\ }\textbf {\bibinfo
  {volume} {80}},\ \bibinfo {pages} {4843} (\bibinfo {year}
  {1998})}\BibitemShut {NoStop}%
\bibitem [{\citenamefont {{Andersson}}\ \emph
  {et~al.}(1999{\natexlab{a}})\citenamefont {{Andersson}}, \citenamefont
  {{Kokkotas}},\ and\ \citenamefont {{Schutz}}}]{AndKokSch99}%
  \BibitemOpen
  \bibfield  {author} {\bibinfo {author} {\bibfnamefont {N.}~\bibnamefont
  {{Andersson}}}, \bibinfo {author} {\bibfnamefont {K.}~\bibnamefont
  {{Kokkotas}}}, \ and\ \bibinfo {author} {\bibfnamefont {B.~F.}\ \bibnamefont
  {{Schutz}}},\ }\href {\doibase 10.1086/306625} {\bibfield  {journal}
  {\bibinfo  {journal} {\apj}\ }\textbf {\bibinfo {volume} {510}},\ \bibinfo
  {pages} {846} (\bibinfo {year} {1999}{\natexlab{a}})}\BibitemShut {NoStop}%
\bibitem [{\citenamefont {{Bildsten}}(1998)}]{Bild99}%
  \BibitemOpen
  \bibfield  {author} {\bibinfo {author} {\bibfnamefont {L.}~\bibnamefont
  {{Bildsten}}},\ }\href {\doibase 10.1086/311440} {\bibfield  {journal}
  {\bibinfo  {journal} {Astrophys. J. Lett.}\ }\textbf {\bibinfo {volume}
  {501}},\ \bibinfo {pages} {L89} (\bibinfo {year} {1998})}\BibitemShut
  {NoStop}%
\bibitem [{\citenamefont {Owen}\ \emph {et~al.}(1998)\citenamefont {Owen},
  \citenamefont {Lindblom}, \citenamefont {Cutler}, \citenamefont {Schutz},
  \citenamefont {Vecchio},\ and\ \citenamefont {Andersson}}]{OLCSVA}%
  \BibitemOpen
  \bibfield  {author} {\bibinfo {author} {\bibfnamefont {B.~J.}\ \bibnamefont
  {Owen}}, \bibinfo {author} {\bibfnamefont {L.}~\bibnamefont {Lindblom}},
  \bibinfo {author} {\bibfnamefont {C.}~\bibnamefont {Cutler}}, \bibinfo
  {author} {\bibfnamefont {B.~F.}\ \bibnamefont {Schutz}}, \bibinfo {author}
  {\bibfnamefont {A.}~\bibnamefont {Vecchio}}, \ and\ \bibinfo {author}
  {\bibfnamefont {N.}~\bibnamefont {Andersson}},\ }\href {\doibase
  10.1103/PhysRevD.58.084020} {\bibfield  {journal} {\bibinfo  {journal} {Phys.
  Rev. D}\ }\textbf {\bibinfo {volume} {58}},\ \bibinfo {pages} {084020}
  (\bibinfo {year} {1998})}\BibitemShut {NoStop}%
\bibitem [{\citenamefont {{Andersson}}\ \emph
  {et~al.}(1999{\natexlab{b}})\citenamefont {{Andersson}}, \citenamefont
  {{Kokkotas}},\ and\ \citenamefont {{Stergioulas}}}]{Andersoon99}%
  \BibitemOpen
  \bibfield  {author} {\bibinfo {author} {\bibfnamefont {N.}~\bibnamefont
  {{Andersson}}}, \bibinfo {author} {\bibfnamefont {K.~D.}\ \bibnamefont
  {{Kokkotas}}}, \ and\ \bibinfo {author} {\bibfnamefont {N.}~\bibnamefont
  {{Stergioulas}}},\ }\href {\doibase 10.1086/307082} {\bibfield  {journal}
  {\bibinfo  {journal} {\apj}\ }\textbf {\bibinfo {volume} {516}},\ \bibinfo
  {pages} {307} (\bibinfo {year} {1999}{\natexlab{b}})}\BibitemShut {NoStop}%
\bibitem [{\citenamefont {{Watts}}(2012)}]{Watts2012}%
  \BibitemOpen
  \bibfield  {author} {\bibinfo {author} {\bibfnamefont {A.~L.}\ \bibnamefont
  {{Watts}}},\ }\href {\doibase 10.1146/annurev-astro-040312-132617} {\bibfield
   {journal} {\bibinfo  {journal} {Ann. Rev. of Astron. and Astrophys.}\
  }\textbf {\bibinfo {volume} {50}},\ \bibinfo {pages} {609} (\bibinfo {year}
  {2012})}\BibitemShut {NoStop}%
\bibitem [{\citenamefont {{Strohmayer}}\ and\ \citenamefont
  {{Mahmoodifar}}(2014{\natexlab{a}})}]{2013r-modeDiscovery}%
  \BibitemOpen
  \bibfield  {author} {\bibinfo {author} {\bibfnamefont {T.}~\bibnamefont
  {{Strohmayer}}}\ and\ \bibinfo {author} {\bibfnamefont {S.}~\bibnamefont
  {{Mahmoodifar}}},\ }\href {\doibase 10.1088/0004-637X/784/1/72} {\bibfield
  {journal} {\bibinfo  {journal} {\apj}\ }\textbf {\bibinfo {volume} {784}},\
  \bibinfo {eid} {72} (\bibinfo {year} {2014}{\natexlab{a}})}\BibitemShut
  {NoStop}%
\bibitem [{\citenamefont {{Strohmayer}}\ and\ \citenamefont
  {{Mahmoodifar}}(2014{\natexlab{b}})}]{r-modeAnotherDiscovery}%
  \BibitemOpen
  \bibfield  {author} {\bibinfo {author} {\bibfnamefont {T.}~\bibnamefont
  {{Strohmayer}}}\ and\ \bibinfo {author} {\bibfnamefont {S.}~\bibnamefont
  {{Mahmoodifar}}},\ }\href {\doibase 10.1088/2041-8205/793/2/L38} {\bibfield
  {journal} {\bibinfo  {journal} {Astrophys. J. Lett.}\ }\textbf {\bibinfo
  {volume} {793}},\ \bibinfo {eid} {L38} (\bibinfo {year}
  {2014}{\natexlab{b}})}\BibitemShut {NoStop}%
\bibitem [{\citenamefont {{Andersson}}\ \emph {et~al.}(2014)\citenamefont
  {{Andersson}}, \citenamefont {{Jones}},\ and\ \citenamefont
  {{Ho}}}]{Andersson2014}%
  \BibitemOpen
  \bibfield  {author} {\bibinfo {author} {\bibfnamefont {N.}~\bibnamefont
  {{Andersson}}}, \bibinfo {author} {\bibfnamefont {D.~I.}\ \bibnamefont
  {{Jones}}}, \ and\ \bibinfo {author} {\bibfnamefont {W.~C.~G.}\ \bibnamefont
  {{Ho}}},\ }\href {\doibase 10.1093/mnras/stu870} {\bibfield  {journal}
  {\bibinfo  {journal} {Mon. Not. R. Astron. Soc.}\ }\textbf {\bibinfo {volume}
  {442}},\ \bibinfo {pages} {1786} (\bibinfo {year} {2014})}\BibitemShut
  {NoStop}%
\bibitem [{\citenamefont {{Lee}}(2014)}]{Lee2014}%
  \BibitemOpen
  \bibfield  {author} {\bibinfo {author} {\bibfnamefont {U.}~\bibnamefont
  {{Lee}}},\ }\href {\doibase 10.1093/mnras/stu1077} {\bibfield  {journal}
  {\bibinfo  {journal} {Mon. Not. R. Astron. Soc.}\ }\textbf {\bibinfo {volume}
  {442}},\ \bibinfo {pages} {3037} (\bibinfo {year} {2014})}\BibitemShut
  {NoStop}%
\bibitem [{\citenamefont {{Provost}}\ \emph {et~al.}(1981)\citenamefont
  {{Provost}}, \citenamefont {{Berthomieu}},\ and\ \citenamefont
  {{Rocca}}}]{Provost1981}%
  \BibitemOpen
  \bibfield  {author} {\bibinfo {author} {\bibfnamefont {J.}~\bibnamefont
  {{Provost}}}, \bibinfo {author} {\bibfnamefont {G.}~\bibnamefont
  {{Berthomieu}}}, \ and\ \bibinfo {author} {\bibfnamefont {A.}~\bibnamefont
  {{Rocca}}},\ }\href@noop {} {\bibfield  {journal} {\bibinfo  {journal}
  {Astron. and Astrophys.}\ }\textbf {\bibinfo {volume} {94}},\ \bibinfo
  {pages} {126} (\bibinfo {year} {1981})}\BibitemShut {NoStop}%
\bibitem [{\citenamefont {Lindblom}\ \emph {et~al.}(1999)\citenamefont
  {Lindblom}, \citenamefont {Mendell},\ and\ \citenamefont
  {Owen}}]{LindOwen99}%
  \BibitemOpen
  \bibfield  {author} {\bibinfo {author} {\bibfnamefont {L.}~\bibnamefont
  {Lindblom}}, \bibinfo {author} {\bibfnamefont {G.}~\bibnamefont {Mendell}}, \
  and\ \bibinfo {author} {\bibfnamefont {B.~J.}\ \bibnamefont {Owen}},\ }\href
  {\doibase 10.1103/PhysRevD.60.064006} {\bibfield  {journal} {\bibinfo
  {journal} {Phys. Rev. D}\ }\textbf {\bibinfo {volume} {60}},\ \bibinfo
  {pages} {064006} (\bibinfo {year} {1999})}\BibitemShut {NoStop}%
\bibitem [{\citenamefont {Lockitch}\ \emph {et~al.}(2003)\citenamefont
  {Lockitch}, \citenamefont {Friedman},\ and\ \citenamefont
  {Andersson}}]{Lock03}%
  \BibitemOpen
  \bibfield  {author} {\bibinfo {author} {\bibfnamefont {K.~H.}\ \bibnamefont
  {Lockitch}}, \bibinfo {author} {\bibfnamefont {J.~L.}\ \bibnamefont
  {Friedman}}, \ and\ \bibinfo {author} {\bibfnamefont {N.}~\bibnamefont
  {Andersson}},\ }\href {\doibase 10.1103/PhysRevD.68.124010} {\bibfield
  {journal} {\bibinfo  {journal} {\prd}\ }\textbf {\bibinfo {volume} {68}},\
  \bibinfo {pages} {124010} (\bibinfo {year} {2003})}\BibitemShut {NoStop}%
\bibitem [{\citenamefont {{Stergioulas}}\ and\ \citenamefont
  {{Font}}(2001)}]{Ster}%
  \BibitemOpen
  \bibfield  {author} {\bibinfo {author} {\bibfnamefont {N.}~\bibnamefont
  {{Stergioulas}}}\ and\ \bibinfo {author} {\bibfnamefont {J.~A.}\ \bibnamefont
  {{Font}}},\ }\href {\doibase 10.1103/PhysRevLett.86.1148} {\bibfield
  {journal} {\bibinfo  {journal} {Phys. Rev. Lett.}\ }\textbf {\bibinfo
  {volume} {86}},\ \bibinfo {pages} {1148} (\bibinfo {year}
  {2001})}\BibitemShut {NoStop}%
\bibitem [{\citenamefont {{Gaertig}}\ and\ \citenamefont
  {{Kokkotas}}(2008)}]{G}%
  \BibitemOpen
  \bibfield  {author} {\bibinfo {author} {\bibfnamefont {E.}~\bibnamefont
  {{Gaertig}}}\ and\ \bibinfo {author} {\bibfnamefont {K.~D.}\ \bibnamefont
  {{Kokkotas}}},\ }\href {\doibase 10.1103/PhysRevD.78.064063} {\bibfield
  {journal} {\bibinfo  {journal} {\prd}\ }\textbf {\bibinfo {volume} {78}},\
  \bibinfo {eid} {064063} (\bibinfo {year} {2008})}\BibitemShut {NoStop}%
\bibitem [{\citenamefont {{Yoshida}}\ \emph {et~al.}(2005)\citenamefont
  {{Yoshida}}, \citenamefont {{Yoshida}},\ and\ \citenamefont
  {{Eriguchi}}}]{YoshidaMain}%
  \BibitemOpen
  \bibfield  {author} {\bibinfo {author} {\bibfnamefont {S.}~\bibnamefont
  {{Yoshida}}}, \bibinfo {author} {\bibfnamefont {S.}~\bibnamefont
  {{Yoshida}}}, \ and\ \bibinfo {author} {\bibfnamefont {Y.}~\bibnamefont
  {{Eriguchi}}},\ }\href {\doibase 10.1111/j.1365-2966.2004.08436.x} {\bibfield
   {journal} {\bibinfo  {journal} {Mon. Not. R. Astron. Soc.}\ }\textbf
  {\bibinfo {volume} {356}},\ \bibinfo {pages} {217} (\bibinfo {year}
  {2005})}\BibitemShut {NoStop}%
\bibitem [{\citenamefont {Read}\ \emph {et~al.}(2009)\citenamefont {Read},
  \citenamefont {Lackey}, \citenamefont {Owen},\ and\ \citenamefont
  {Friedman}}]{Read4P}%
  \BibitemOpen
  \bibfield  {author} {\bibinfo {author} {\bibfnamefont {J.~S.}\ \bibnamefont
  {Read}}, \bibinfo {author} {\bibfnamefont {B.~D.}\ \bibnamefont {Lackey}},
  \bibinfo {author} {\bibfnamefont {B.~J.}\ \bibnamefont {Owen}}, \ and\
  \bibinfo {author} {\bibfnamefont {J.~L.}\ \bibnamefont {Friedman}},\ }\href
  {\doibase 10.1103/PhysRevD.79.124032} {\bibfield  {journal} {\bibinfo
  {journal} {\prd}\ }\textbf {\bibinfo {volume} {79}},\ \bibinfo {eid} {124032}
  (\bibinfo {year} {2009})}\BibitemShut {NoStop}%
\bibitem [{\citenamefont {Lindblom}\ and\ \citenamefont
  {Indik}(2012)}]{LindIndik}%
  \BibitemOpen
  \bibfield  {author} {\bibinfo {author} {\bibfnamefont {L.}~\bibnamefont
  {Lindblom}}\ and\ \bibinfo {author} {\bibfnamefont {N.~M.}\ \bibnamefont
  {Indik}},\ }\href {\doibase 10.1103/PhysRevD.86.084003} {\bibfield  {journal}
  {\bibinfo  {journal} {\prd}\ }\textbf {\bibinfo {volume} {D86}},\ \bibinfo
  {pages} {084003} (\bibinfo {year} {2012})}\BibitemShut {NoStop}%
\bibitem [{\citenamefont {{Demorest}}\ \emph {et~al.}(2010)\citenamefont
  {{Demorest}}, \citenamefont {{Pennucci}}, \citenamefont {{Ransom}},
  \citenamefont {{Roberts}},\ and\ \citenamefont {{Hessels}}}]{Demorest}%
  \BibitemOpen
  \bibfield  {author} {\bibinfo {author} {\bibfnamefont {P.~B.}\ \bibnamefont
  {{Demorest}}}, \bibinfo {author} {\bibfnamefont {T.}~\bibnamefont
  {{Pennucci}}}, \bibinfo {author} {\bibfnamefont {S.~M.}\ \bibnamefont
  {{Ransom}}}, \bibinfo {author} {\bibfnamefont {M.~S.~E.}\ \bibnamefont
  {{Roberts}}}, \ and\ \bibinfo {author} {\bibfnamefont {J.~W.~T.}\
  \bibnamefont {{Hessels}}},\ }\href {\doibase 10.1038/nature09466} {\bibfield
  {journal} {\bibinfo  {journal} {\nat}\ }\textbf {\bibinfo {volume} {467}},\
  \bibinfo {pages} {1081} (\bibinfo {year} {2010})}\BibitemShut {NoStop}%
\bibitem [{\citenamefont {{Abbott}}\ and\ \citenamefont
  {et~al.}(2008)}]{Crab2008}%
  \BibitemOpen
  \bibfield  {author} {\bibinfo {author} {\bibfnamefont {B.}~\bibnamefont
  {{Abbott}}}\ and\ \bibinfo {author} {\bibnamefont {et~al.}},\ }\href
  {\doibase 10.1086/591526} {\bibfield  {journal} {\bibinfo  {journal} {\apj}\
  }\textbf {\bibinfo {volume} {683}},\ \bibinfo {pages} {L45} (\bibinfo {year}
  {2008})}\BibitemShut {NoStop}%
\bibitem [{\citenamefont {{Astone}}\ \emph {et~al.}(2014)\citenamefont
  {{Astone}}, \citenamefont {{Colla}}, \citenamefont {{D'Antonio}},
  \citenamefont {{Frasca}}, \citenamefont {{Palomba}},\ and\ \citenamefont
  {{Serafinelli}}}]{narrowband}%
  \BibitemOpen
  \bibfield  {author} {\bibinfo {author} {\bibfnamefont {P.}~\bibnamefont
  {{Astone}}}, \bibinfo {author} {\bibfnamefont {A.}~\bibnamefont {{Colla}}},
  \bibinfo {author} {\bibfnamefont {S.}~\bibnamefont {{D'Antonio}}}, \bibinfo
  {author} {\bibfnamefont {S.}~\bibnamefont {{Frasca}}}, \bibinfo {author}
  {\bibfnamefont {C.}~\bibnamefont {{Palomba}}}, \ and\ \bibinfo {author}
  {\bibfnamefont {R.}~\bibnamefont {{Serafinelli}}},\ }\href {\doibase
  10.1103/PhysRevD.89.062008} {\bibfield  {journal} {\bibinfo  {journal}
  {\prd}\ }\textbf {\bibinfo {volume} {89}},\ \bibinfo {eid} {062008} (\bibinfo
  {year} {2014})}\BibitemShut {NoStop}%
\bibitem [{\citenamefont {{Lattimer}}\ and\ \citenamefont
  {{Prakash}}(2011)}]{lp10}%
  \BibitemOpen
  \bibfield  {author} {\bibinfo {author} {\bibfnamefont {J.~M.}\ \bibnamefont
  {{Lattimer}}}\ and\ \bibinfo {author} {\bibfnamefont {M.}~\bibnamefont
  {{Prakash}}},\ }in\ \href@noop {} {\emph {\bibinfo {booktitle} {{From Nuclei
  to Stars}}}},\ \bibinfo {editor} {edited by\ \bibinfo {editor} {\bibfnamefont
  {S.}~\bibnamefont {Lee}}}\ (\bibinfo  {publisher} {{World Scientific}},\
  \bibinfo {address} {Singapore},\ \bibinfo {year} {2011})\ p.\ \bibinfo
  {pages} {275}\BibitemShut {NoStop}%
\bibitem [{\citenamefont {{Lattimer}}\ and\ \citenamefont
  {{Prakash}}(2007)}]{lp07}%
  \BibitemOpen
  \bibfield  {author} {\bibinfo {author} {\bibfnamefont {J.~M.}\ \bibnamefont
  {{Lattimer}}}\ and\ \bibinfo {author} {\bibfnamefont {M.}~\bibnamefont
  {{Prakash}}},\ }\href {\doibase 10.1016/j.physrep.2007.02.003} {\bibfield
  {journal} {\bibinfo  {journal} {Phys. Rep.}\ }\textbf {\bibinfo {volume}
  {442}},\ \bibinfo {pages} {109} (\bibinfo {year} {2007})}\BibitemShut
  {NoStop}%
\bibitem [{\citenamefont {{Manchester}}\ \emph {et~al.}(2005)\citenamefont
  {{Manchester}}, \citenamefont {{Hobbs}}, \citenamefont {{Teoh}},\ and\
  \citenamefont {{Hobbs}}}]{ATNF}%
  \BibitemOpen
  \bibfield  {author} {\bibinfo {author} {\bibfnamefont {R.~N.}\ \bibnamefont
  {{Manchester}}}, \bibinfo {author} {\bibfnamefont {G.~B.}\ \bibnamefont
  {{Hobbs}}}, \bibinfo {author} {\bibfnamefont {A.}~\bibnamefont {{Teoh}}}, \
  and\ \bibinfo {author} {\bibfnamefont {M.}~\bibnamefont {{Hobbs}}},\ }\href
  {\doibase 10.1086/428488} {\bibfield  {journal} {\bibinfo  {journal} {Astron.
  J.}\ }\textbf {\bibinfo {volume} {129}},\ \bibinfo {pages} {1993} (\bibinfo
  {year} {2005})}\BibitemShut {NoStop}%
\bibitem [{\citenamefont {{Bildsten}}\ and\ \citenamefont
  {{Ushomirsky}}(2000)}]{BilUsh2000}%
  \BibitemOpen
  \bibfield  {author} {\bibinfo {author} {\bibfnamefont {L.}~\bibnamefont
  {{Bildsten}}}\ and\ \bibinfo {author} {\bibfnamefont {G.}~\bibnamefont
  {{Ushomirsky}}},\ }\href {\doibase 10.1086/312454} {\bibfield  {journal}
  {\bibinfo  {journal} {Astrophys. J. Lett.}\ }\textbf {\bibinfo {volume}
  {529}},\ \bibinfo {pages} {L33} (\bibinfo {year} {2000})}\BibitemShut
  {NoStop}%
\bibitem [{\citenamefont {{Levin}}\ and\ \citenamefont
  {{Ushomirsky}}(2001)}]{LevUsh2001}%
  \BibitemOpen
  \bibfield  {author} {\bibinfo {author} {\bibfnamefont {Y.}~\bibnamefont
  {{Levin}}}\ and\ \bibinfo {author} {\bibfnamefont {G.}~\bibnamefont
  {{Ushomirsky}}},\ }\href {\doibase 10.1046/j.1365-8711.2001.04323.x}
  {\bibfield  {journal} {\bibinfo  {journal} {Mon. Not. R. Astron. Soc.}\
  }\textbf {\bibinfo {volume} {324}},\ \bibinfo {pages} {917} (\bibinfo {year}
  {2001})}\BibitemShut {NoStop}%
\bibitem [{\citenamefont {{Yoshida}}\ and\ \citenamefont
  {{Lee}}(2000)}]{YosLee2000}%
  \BibitemOpen
  \bibfield  {author} {\bibinfo {author} {\bibfnamefont {S.}~\bibnamefont
  {{Yoshida}}}\ and\ \bibinfo {author} {\bibfnamefont {U.}~\bibnamefont
  {{Lee}}},\ }\href {\doibase 10.1086/313410} {\bibfield  {journal} {\bibinfo
  {journal} {Astrophys. J. Suppl.}\ }\textbf {\bibinfo {volume} {129}},\
  \bibinfo {pages} {353} (\bibinfo {year} {2000})}\BibitemShut {NoStop}%
\bibitem [{\citenamefont {{Passamonti}}\ \emph {et~al.}(2009)\citenamefont
  {{Passamonti}}, \citenamefont {{Haskell}}, \citenamefont {{Andersson}},
  \citenamefont {{Jones}},\ and\ \citenamefont {{Hawke}}}]{PasHas2009}%
  \BibitemOpen
  \bibfield  {author} {\bibinfo {author} {\bibfnamefont {A.}~\bibnamefont
  {{Passamonti}}}, \bibinfo {author} {\bibfnamefont {B.}~\bibnamefont
  {{Haskell}}}, \bibinfo {author} {\bibfnamefont {N.}~\bibnamefont
  {{Andersson}}}, \bibinfo {author} {\bibfnamefont {D.~I.}\ \bibnamefont
  {{Jones}}}, \ and\ \bibinfo {author} {\bibfnamefont {I.}~\bibnamefont
  {{Hawke}}},\ }\href {\doibase 10.1111/j.1365-2966.2009.14408.x} {\bibfield
  {journal} {\bibinfo  {journal} {Mon. Not. R. Astron. Soc.}\ }\textbf
  {\bibinfo {volume} {394}},\ \bibinfo {pages} {730} (\bibinfo {year}
  {2009})}\BibitemShut {NoStop}%
\bibitem [{\citenamefont {{Ho}}\ and\ \citenamefont {{Lai}}(2000)}]{HoLai2000}%
  \BibitemOpen
  \bibfield  {author} {\bibinfo {author} {\bibfnamefont {W.~C.~G.}\
  \bibnamefont {{Ho}}}\ and\ \bibinfo {author} {\bibfnamefont {D.}~\bibnamefont
  {{Lai}}},\ }\href {\doibase 10.1086/317085} {\bibfield  {journal} {\bibinfo
  {journal} {\apj}\ }\textbf {\bibinfo {volume} {543}},\ \bibinfo {pages} {386}
  (\bibinfo {year} {2000})}\BibitemShut {NoStop}%
\bibitem [{\citenamefont {{Morsink}}\ and\ \citenamefont
  {{Rezania}}(2002)}]{MorRez2002}%
  \BibitemOpen
  \bibfield  {author} {\bibinfo {author} {\bibfnamefont {S.~M.}\ \bibnamefont
  {{Morsink}}}\ and\ \bibinfo {author} {\bibfnamefont {V.}~\bibnamefont
  {{Rezania}}},\ }\href {\doibase 10.1086/341190} {\bibfield  {journal}
  {\bibinfo  {journal} {\apj}\ }\textbf {\bibinfo {volume} {574}},\ \bibinfo
  {pages} {908} (\bibinfo {year} {2002})}\BibitemShut {NoStop}%
\bibitem [{\citenamefont {{Lander}}\ \emph {et~al.}(2010)\citenamefont
  {{Lander}}, \citenamefont {{Jones}},\ and\ \citenamefont
  {{Passamonti}}}]{LanJones2010}%
  \BibitemOpen
  \bibfield  {author} {\bibinfo {author} {\bibfnamefont {S.~K.}\ \bibnamefont
  {{Lander}}}, \bibinfo {author} {\bibfnamefont {D.~I.}\ \bibnamefont
  {{Jones}}}, \ and\ \bibinfo {author} {\bibfnamefont {A.}~\bibnamefont
  {{Passamonti}}},\ }\href {\doibase 10.1111/j.1365-2966.2010.16435.x}
  {\bibfield  {journal} {\bibinfo  {journal} {Mon. Not. R. Astron. Soc.}\
  }\textbf {\bibinfo {volume} {405}},\ \bibinfo {pages} {318} (\bibinfo {year}
  {2010})}\BibitemShut {NoStop}%
\bibitem [{\citenamefont {{Lockitch}}\ \emph {et~al.}(2001)\citenamefont
  {{Lockitch}}, \citenamefont {{Andersson}},\ and\ \citenamefont
  {{Friedman}}}]{Lock01}%
  \BibitemOpen
  \bibfield  {author} {\bibinfo {author} {\bibfnamefont {K.~H.}\ \bibnamefont
  {{Lockitch}}}, \bibinfo {author} {\bibfnamefont {N.}~\bibnamefont
  {{Andersson}}}, \ and\ \bibinfo {author} {\bibfnamefont {J.~L.}\ \bibnamefont
  {{Friedman}}},\ }\href {\doibase 10.1103/PhysRevD.63.024019} {\bibfield
  {journal} {\bibinfo  {journal} {\prd}\ }\textbf {\bibinfo {volume} {63}},\
  \bibinfo {eid} {024019} (\bibinfo {year} {2001})}\BibitemShut {NoStop}%
\bibitem [{\citenamefont {Carroll}(2004)}]{carroll2004}%
  \BibitemOpen
  \bibfield  {author} {\bibinfo {author} {\bibfnamefont {S.}~\bibnamefont
  {Carroll}},\ }\href {http://books.google.com/books?id=1SKFQgAACAAJ} {\emph
  {\bibinfo {title} {Spacetime and Geometry: An Introduction to General
  Relativity}}}\ (\bibinfo  {publisher} {Addison-Wesley Longman, San
  Francisco},\ \bibinfo {year} {2004})\BibitemShut {NoStop}%
\bibitem [{\citenamefont {{Lindblom}}(1992)}]{Lindblom1992}%
  \BibitemOpen
  \bibfield  {author} {\bibinfo {author} {\bibfnamefont {L.}~\bibnamefont
  {{Lindblom}}},\ }\href {\doibase 10.1086/171882} {\bibfield  {journal}
  {\bibinfo  {journal} {\apj}\ }\textbf {\bibinfo {volume} {398}},\ \bibinfo
  {pages} {569} (\bibinfo {year} {1992})}\BibitemShut {NoStop}%
\bibitem [{\citenamefont {{Lockitch}}(1999)}]{LockThesis}%
  \BibitemOpen
  \bibfield  {author} {\bibinfo {author} {\bibfnamefont {K.~H.}\ \bibnamefont
  {{Lockitch}}},\ }\emph {\bibinfo {title} {{Stability and rotational mixing of
  modes in Newtonian and relativistic stars}}},\ \href@noop {} {Ph.D. thesis},\
  \bibinfo  {school} {THE UNIVERSITY OF WISCONSIN - MILWAUKEE} (\bibinfo {year}
  {1999})\BibitemShut {NoStop}%
\bibitem [{\citenamefont {{Haensel}}\ and\ \citenamefont
  {{Proszynski}}(1982)}]{HP}%
  \BibitemOpen
  \bibfield  {author} {\bibinfo {author} {\bibfnamefont {P.}~\bibnamefont
  {{Haensel}}}\ and\ \bibinfo {author} {\bibfnamefont {M.}~\bibnamefont
  {{Proszynski}}},\ }\href {\doibase 10.1086/160080} {\bibfield  {journal}
  {\bibinfo  {journal} {\apj}\ }\textbf {\bibinfo {volume} {258}},\ \bibinfo
  {pages} {306} (\bibinfo {year} {1982})}\BibitemShut {NoStop}%
\bibitem [{\citenamefont {Press}(2002)}]{NumRec}%
  \BibitemOpen
  \bibfield  {author} {\bibinfo {author} {\bibfnamefont {W.}~\bibnamefont
  {Press}},\ }\href {http://books.google.com/books?id=mSLhDt\_XIUQC} {\emph
  {\bibinfo {title} {Numerical Recipes in C++: The Art of Scientific
  Computing}}}\ (\bibinfo  {publisher} {Cambridge University Press, New York},\
  \bibinfo {year} {2002})\BibitemShut {NoStop}%
\bibitem [{\citenamefont {{Hartle}}(1967)}]{Hartle67}%
  \BibitemOpen
  \bibfield  {author} {\bibinfo {author} {\bibfnamefont {J.~B.}\ \bibnamefont
  {{Hartle}}},\ }\href {\doibase 10.1086/149400} {\bibfield  {journal}
  {\bibinfo  {journal} {\apj}\ }\textbf {\bibinfo {volume} {150}},\ \bibinfo
  {pages} {1005} (\bibinfo {year} {1967})}\BibitemShut {NoStop}%
\bibitem [{\citenamefont {Arfken}\ \emph {et~al.}(2005)\citenamefont {Arfken},
  \citenamefont {Weber},\ and\ \citenamefont {Harris}}]{arfken2005}%
  \BibitemOpen
  \bibfield  {author} {\bibinfo {author} {\bibfnamefont {G.}~\bibnamefont
  {Arfken}}, \bibinfo {author} {\bibfnamefont {H.}~\bibnamefont {Weber}}, \
  and\ \bibinfo {author} {\bibfnamefont {F.}~\bibnamefont {Harris}},\ }\href
  {http://books.google.com/books?id=tNtijk2iBSMC} {\emph {\bibinfo {title}
  {Mathematical Methods For Physicists International Student Edition}}}\
  (\bibinfo  {publisher} {Elsevier Science, Oxford},\ \bibinfo {year}
  {2005})\BibitemShut {NoStop}%
\bibitem [{\citenamefont {Mohr}\ \emph {et~al.}(2012)\citenamefont {Mohr},
  \citenamefont {Taylor},\ and\ \citenamefont {Newell}}]{Gnow}%
  \BibitemOpen
  \bibfield  {author} {\bibinfo {author} {\bibfnamefont {P.~J.}\ \bibnamefont
  {Mohr}}, \bibinfo {author} {\bibfnamefont {B.~N.}\ \bibnamefont {Taylor}}, \
  and\ \bibinfo {author} {\bibfnamefont {D.~B.}\ \bibnamefont {Newell}},\
  }\href {\doibase 10.1103/RevModPhys.84.1527} {\bibfield  {journal} {\bibinfo
  {journal} {Rev. Mod. Phys.}\ }\textbf {\bibinfo {volume} {84}},\ \bibinfo
  {pages} {1527} (\bibinfo {year} {2012})}\BibitemShut {NoStop}%
\bibitem [{\citenamefont {Mohr}\ and\ \citenamefont {Taylor}(2000)}]{Gthen}%
  \BibitemOpen
  \bibfield  {author} {\bibinfo {author} {\bibfnamefont {P.}~\bibnamefont
  {Mohr}}\ and\ \bibinfo {author} {\bibfnamefont {B.}~\bibnamefont {Taylor}},\
  }\href {\doibase 10.1103/RevModPhys.72.351} {\bibfield  {journal} {\bibinfo
  {journal} {Rev. Mod. Phys.}\ }\textbf {\bibinfo {volume} {72}},\ \bibinfo
  {pages} {351} (\bibinfo {year} {2000})}\BibitemShut {NoStop}%
\bibitem [{\citenamefont {Kutner}(2005)}]{kutner2005applied}%
  \BibitemOpen
  \bibfield  {author} {\bibinfo {author} {\bibfnamefont {M.}~\bibnamefont
  {Kutner}},\ }\href@noop {} {\emph {\bibinfo {title} {Applied Linear
  Statistical Models}}},\ McGraw-Hill/Irwin series\ (\bibinfo  {publisher}
  {McGraw-Hill Irwin, New York},\ \bibinfo {year} {2005})\BibitemShut {NoStop}%
\bibitem [{\citenamefont {{Douchin}}\ and\ \citenamefont
  {{Haensel}}(2001)}]{Douchin}%
  \BibitemOpen
  \bibfield  {author} {\bibinfo {author} {\bibfnamefont {F.}~\bibnamefont
  {{Douchin}}}\ and\ \bibinfo {author} {\bibfnamefont {P.}~\bibnamefont
  {{Haensel}}},\ }\href {\doibase 10.1051/0004-6361:20011402} {\bibfield
  {journal} {\bibinfo  {journal} {Astron. and Astrophys.}\ }\textbf {\bibinfo
  {volume} {380}},\ \bibinfo {pages} {151} (\bibinfo {year}
  {2001})}\BibitemShut {NoStop}%
\bibitem [{\citenamefont {Akmal}\ \emph {et~al.}(1998)\citenamefont {Akmal},
  \citenamefont {Pandharipande},\ and\ \citenamefont {Ravenhall}}]{APR}%
  \BibitemOpen
  \bibfield  {author} {\bibinfo {author} {\bibfnamefont {A.}~\bibnamefont
  {Akmal}}, \bibinfo {author} {\bibfnamefont {V.~R.}\ \bibnamefont
  {Pandharipande}}, \ and\ \bibinfo {author} {\bibfnamefont {D.~G.}\
  \bibnamefont {Ravenhall}},\ }\href {\doibase 10.1103/PhysRevC.58.1804}
  {\bibfield  {journal} {\bibinfo  {journal} {Phys. Rev. C}\ }\textbf {\bibinfo
  {volume} {58}},\ \bibinfo {pages} {1804} (\bibinfo {year}
  {1998})}\BibitemShut {NoStop}%
\end{thebibliography}%

\end{document}